\documentclass[preprint,11pt]{elsarticle}
\usepackage[utf8]{inputenc}
\usepackage{amsmath}
\usepackage{amsfonts}
\usepackage{amssymb}
\usepackage{graphicx}
\usepackage{layouts}
\usepackage{algorithm}
\usepackage{algpseudocode}
\usepackage[dvipsnames]{xcolor}
\usepackage{comment}
\usepackage[margin=3cm]{geometry}
\usepackage[page,toc,titletoc,title]{appendix}
\usepackage{amsthm}
\usepackage{tcolorbox}
\usepackage{printlen}
\usepackage{mathtools}
\usepackage{graphbox}
\tcbuselibrary{breakable}
\usepackage{tikz}
\usepackage{amsthm}
\usepackage{lipsum}
\usepackage{natbib}
\usepackage{hyperref}
\theoremstyle{definition}

\journal{Computer Methods in Applied Mechanics and Engineering}

\biboptions{sort&compress}

\begin{document}

\begin{frontmatter}



\title{Tensor network compression using fluid dynamics as a testbed: Analytical foundations in one dimension}

\author{Matthew D. Horner}
\author{Callum W. Duncan}
\affiliation{Aegiq Ltd., Cooper Buildings, Sheffield S1 2NS, United Kingdom}
\author{Oliver T. Brown}
\affiliation{EPCC, University of Edinburgh, Edinburgh, EH8 9BT, United Kingdom}
\author{Stephen M. de Bruyn Kops}
\affiliation{Department of Mechanical and Industrial Engineering, University of Massachusetts Amherst, Amherst, MA 01003, USA}
\author{Muralikrishnan Gopalakrishnan Meena} 
\affiliation{National Center for Computational Sciences, Oak Ridge National Laboratory, Oak Ridge, TN 37831, USA}


\begin{abstract}
High performance computers produce extreme-scale data sets that require sampling or compression if they are to be used to their full potential.  Existing data compression techniques typically exploit features such as sparsity in the data, homogeneity in the data, or {\it a priori} knowledge of what subsets of data are of most interest.  Fluid dynamics data in general do not exhibit these features and so are attractive test beds for generic compression techniques that are objective, robust, and tuneable with respect to information lost due to compression.  Presented here is a method based on tensor networks, specifically matrix product states or tensor trains, that meets these requirements. The method is demonstrated for compression in one-dimension and is extensible to higher dimensionality.  Lossless compression is demonstrated for random Fourier series for sufficiently high bond dimension of the tensor network, with the memory required to store the tensor network scaling directly proportional to the bond dimension. The lossy compression exhibited at lower bond dimension can be well within the relative error of many fluid simulations. The compression algorithm is tested for the time evolution of Burger's equation with excellent results. We additionally demonstrate the capability to perform computations in the compressed form through a tensor network periodic convolution that can be orders of magnitude faster than using fast Fourier transforms and the convolution theorem. In addition to being an attractive method for working with data sets generated by existing computers, the tensor network methods utilised are directly translatable to the emerging paradigm of quantum computing.
\footnote{\textbf{Notice:} This manuscript has been authored in-part by UT-Battelle, LLC, under contract DE-AC05-00OR22725 with the US Department of Energy (DOE). The US government retains and the publisher, by accepting the article for publication, acknowledges that the US government retains a nonexclusive, paid-up, irrevocable, worldwide license to publish or reproduce the published form of this manuscript, or allow others to do so, for US government purposes. DOE will provide public access to these results of federally sponsored research in accordance with the DOE Public Access Plan (\href{https://www.energy.gov/doe-public-access-plan}{https://www.energy.gov/doe-public-access-plan}).}
\end{abstract}




\end{frontmatter}

\tableofcontents

\section{Introduction}

Numerically simulating fluid dynamics problems is one of the most challenging problems in computational science, owing to the multiscale, multiphysics behavior of flows found in nature and engineering applications, which are predominantly turbulent. The extreme scales of phenomena occurring in these problems result in massive spatial and temporal resolution requirements of the simulations, necessitating exascale-class supercomputing architectures  \cite{de2019effects,norman2021unprecedented,atchley2023frontier,riley2023effect,yeung2025gpu,yeung2025small,wilfong2025simulating,bhattacharjee26}. The size of the data output by these simulations currently ranges from 100--300\,terabytes per snapshot. As modern high performance computing (HPC) architectures increase compute capability by orders of magnitude roughly every 3--5 years \cite{vazhkudai2018design,atchley2023frontier,allcock2025aurora}, the spatial and temporal resolutions that these simulations can traverse will increase, resulting in prohibitively large data storage requirements. In this context, effective data reduction and compression techniques play a pivotal role.

Current data compression techniques are predominantly centered around sampling-based approaches and those that exploit sparsity in the data. Sampling techniques for fluid dynamics utilize the inherent structure of the flow fields to compress it into candidate data points that are representative of the global spatial and temporal behavior of the flow \cite{manohar2018data,brewer2023entropy,brewer2025intelligent}. Furthermore, reduced-order modeling (ROM) techniques---such as proper orthogonal decomposition (POD), dynamic mode decomposition (DMD), graph theoretic approaches, machine learning approaches, and quantum computing based approaches---have been demonstrated to effectively extract the predominant features in complex fluid flows \cite{lumley1967structure,schmid2010dynamic,taira2017modal,asztalos2024reduced,gopalakrishnan2018network,bai2019randomized,csala2022comparing}. 

\begin{figure}
\begin{center}
\includegraphics[width=0.98\textwidth]{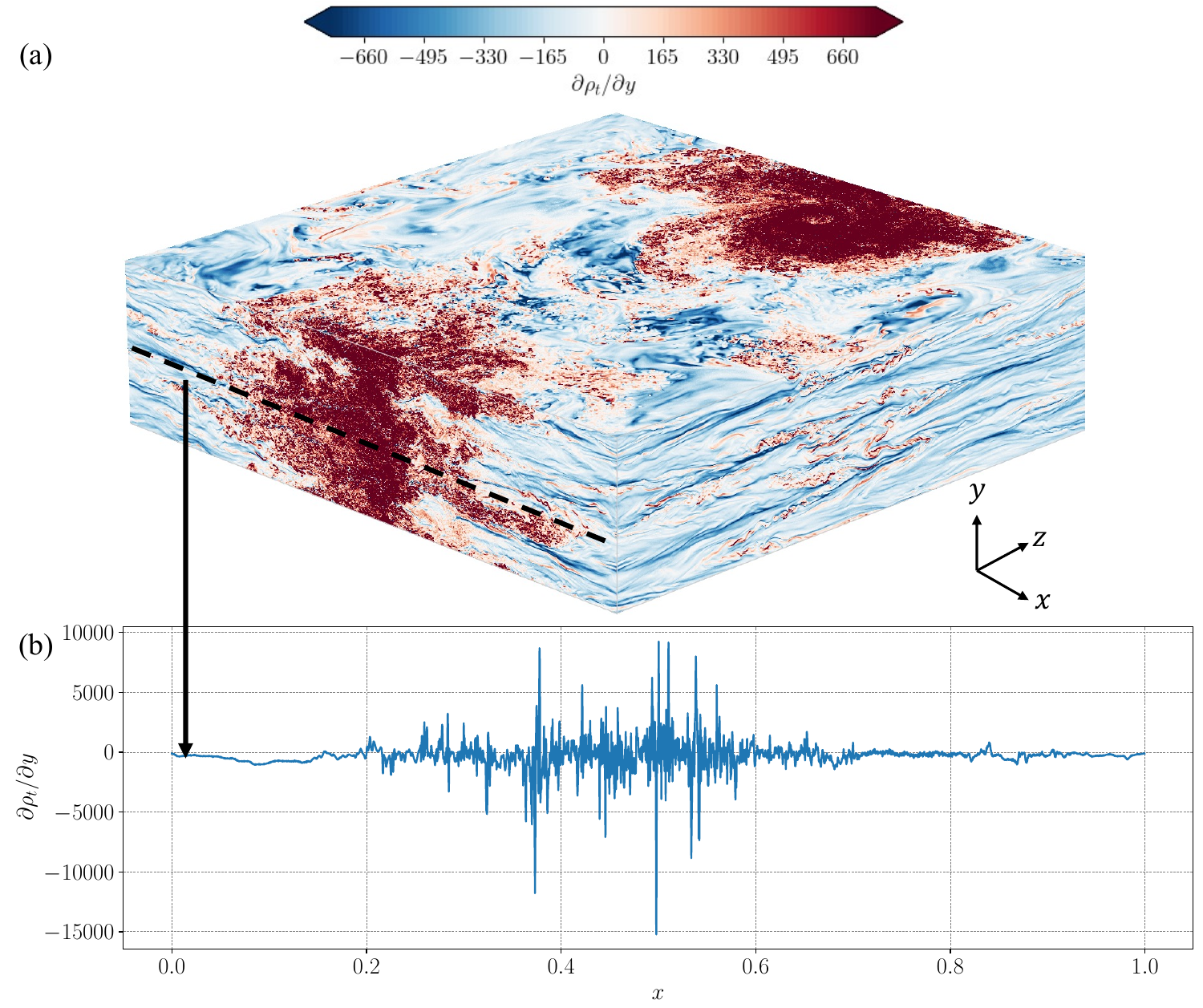}
\end{center}
\caption{(a) A sample flow field from a highly non-isotropic turbulent flow: a stably stratified turbulent flow simulation at Prandtl number 1, Froude number 1, and buoyancy Reynolds number 50 \cite{de2019effects}. Flow field shown is the vertical gradient of total density, $\partial \rho_t / \partial y$. The data volume for each variable in the simulation at a single snapshot is approximately 275 gigabytes. (b) One-dimensional data along a representative slice, depicting the highly non-isotropic nature of the flow.
\label{fig:flow_non-iso-sst}}
\end{figure}

Despite these innovations, important challenges remain. Sampling based methods rely on prior knowledge of the flow field structure to guide the selection of representative data points. ROMs often rely on linearity assumptions and tend to be biased towards large, energy-dominant scales, making them unreliable for accurately capturing small-scale, intermittent features that are central to understanding and modeling turbulent flows. An example of a highly non-isotropic turbulent flow field that poses such challenges for compression and representative sampling is shown in Fig.~\ref{fig:flow_non-iso-sst}. Furthermore, their data storage requirements scale unfavorably to capture complex problems: for a simulation with $L$ grid points, the full data volume grows as $L$ and the parameters and operations to perform the data reduction grow at higher order polynomial complexity.

The importance of length scales extends beyond fluid dynamics. In high energy physics and quantum many-body physics, length scales are a central characteristic of models too, whereby long-scale physics is insensitive to microscopic details, much like the energy cascade of fluid dynamics. Physicists have developed a framework of techniques know as the \textit{renormalization group} to produce effective descriptions of the relevant physics at different length scales, by retaining the degrees of freedom that are important at that particular scale~\cite{RevModPhys.55.583}. In quantum many-body physics for example, typically the relevant physics has low entanglement between widely separated length scales. The density matrix renormalization group (DMRG) and tensor network methods were developed and successfully applied to a range of these many-body systems which would otherwise be numerically intractable with direct numerical simulation (DNS)~\cite{PhysRevLett.69.2863, tensor_networks, PhysRevB.48.10345}.

The ability of tensor networks and DMRG to efficiently describe systems with low correlation between widely-separated length scales suggests application outside of quantum physics. Indeed, recently tensor network methods have emerged as an effective tool to simulate partial differential equations and compress fluid flow data \cite{khoromskij2010quantics,khoromskij2011d,lubasch2018multigrid,garcia2021quantum,pool2024nonlinear,gopalakrishnan2024towards,connor2025tensor-GPE,kiffner2023tensor,connor2025tensor-plasma,ghahremani2024deim}. In the context of turbulent flows specifically, Gourianov et al.~\cite{gourianov2022quantum} demonstrated that matrix product state (MPS) decompositions can be used to substantially reduce the number of parameters required to encode turbulence data relative to a DNS. Subsequent work has utilized tensor networks to consider extensions to plasma dynamics \cite{connor2025tensor-plasma,ye2022quantum}, GPU-accelerated simulation of two-dimensional turbulence \cite{holscher2025quantum}, computing probability distributions of turbulent flows \cite{gourianov2025tensor}, and provide a comprehensive validation of tensor trains to compress three-dimensional incompressible turbulent flows at high Reynolds numbers \cite{pisoni2025compression}.

Tensor networks provide an effective solution for the resolution-scaling bottleneck, but a rigorous analytical foundation---characterizing precisely \emph{when} and \emph{how much} compression is possible, and what operations can be efficiently performed within the compressed representation---has so far been lacking for fluid dynamics data.

In this work, we establish such a foundation for one-dimensional (1D) fluid flow data. We explore the procedure of using matrix product states (MPS) to exactly represent and compress a tensor encoding a discretized representation of a continuous function, which represents an $L$ grid point data structure, into a $\log(L)$-site MPS. This encoding is of interest not only as a classical compression scheme, but also because MPS states are a natural representation for quantum computing hardware, providing a direct pathway to port fluid-dynamics algorithms and data to quantum devices. We derive analytical bounds on the compression complexity of this representation with respect to the maximum bond dimension of the MPS. We demonstrate that for a function represented as Fourier modes, lossless compression is guaranteed whenever the number of active modes is below a threshold set by the grid size. We also demonstrate how operations like nonlinear convolutions can be implemented on the data while compressed, demonstrating computational speed up compared to classical convolution operation beyond a certain grid size.

The rest of the paper is structured as follows. Section~\ref{sec:mps_encoding} reviews the theoretical background for encoding a function into a tensor, truncation strategies to compress the MPS, and the computational complexity of the compression method. In Section~\ref{sec:1d-fourier}, we analyze the MPS compression of 1D Fourier series, covering two distinct encoding approaches, the Nyquist limit on representable frequencies, and inexact (lossy) compression. Section~\ref{sec:application-1d-burgers} demonstrates the application of this framework on a 1D fluid dynamics problem -- the Burgers' equation. We show compression of the velocity field, analyze the evolution of Fourier modes in time dynamics, and demonstrate the effectiveness of tensor-network-based convolution to perform the nonlinear operation. Finally, Section~\ref{sec:discussion-conclusion} provides discussion and concluding remarks.

\section{Theoretical background \label{sec:mps_encoding}}

\subsection{Encoding a scalar function into a matrix product state \label{sec:background}}
To encode a one-dimensional function $f(x)$ on a finite domain into a MPS, we first discretise the domain of the function into $2^N$ equally-spaced points $x_n$, where $n = 1,2,\ldots,2^N$. We then approximate the function by defining the vector $
f_n = f(x_n)$ whose values are the function sampled on the discrete lattice points. Instead of labelling our lattice sites with a single index $n$, we label them with an $N$-bit string $\sigma = (\sigma_1,\sigma_2,\ldots,\sigma_N)$. Each bit string corresponds to the lattice site with position
\begin{equation}
x(\sigma) = \sum_{j = 1}^N \frac{\sigma_j}{2^j}, \label{eq:binary_to_decimal}
\end{equation}
where $x(\sigma)$ is the map between the bit string $\sigma$ and the position $x_n$ it encodes, and each bit $\sigma_j$ corresponds to a different length scale, where the $j$th bit corresponds to the length scale $1/2^j$. We then introduce an $N$th order tensor $T^{\sigma_1 \sigma_2 \ldots \sigma_N }$, where $\sigma_i \in \{0,1\}$ are referred to as physical indices, whose components are defined as the components of the function sampled on the lattice as
\begin{equation}
T^{\sigma_1 \sigma_2 \ldots \sigma_N } = f(x(\sigma)).
\end{equation}
This process is a simple reshaping of the $2^N$ components $f_n$ into an $N$th order tensor with two-dimensional indices. For this reason, this is an exact representation of the function and is lossless.

We can then use successive singular value decompositions (SVD) to decompose this rank-$N$ tensor into a product of $N$ smaller tensors. To do this, we first reshape it into a rectangular matrix and then apply the SVD in the following way:
\begin{align}
T^{\sigma_1 \sigma_2 \ldots \sigma_N} & \equiv T^{\sigma_1 ( \sigma_2 \ldots \sigma_N)}  \\
& = (U \Sigma V^\dagger)^{\sigma_1 (\sigma_2 \ldots \sigma_N)}  \\
& = \sum_{a_1 = 1}^{\chi_1} U^{\sigma_1}_{a_1} s_{a_1} (V^\dagger)_{a_1}^{  (\sigma_2 \ldots \sigma_N)} \\
& \equiv \sum_{a_1 = 1}^{\chi_1}  (A_1)^{\sigma_1}_{a_1} T^{\sigma_2 \ldots \sigma_N}_{a_1} 
\end{align}
where $\chi_1 = \min \{ 2, 2^{N-1} \} = 2$ is the number of singular values for this particular SVD, $\{s_{a_i} \}$ are the singular values, and brackets around multiple indices denote a multi-index. In the final line we define $(A_1)^{\sigma_1}_{a_1} = U^{\sigma_1}_{a_1}$ and absorb the other terms into a single tensor $T^{\sigma_2 \ldots \sigma_N}_{a_1}$. For the $n$th iteration, we apply the above steps to the remaining tensor
\begin{equation}
T_{a_{n-1}}^{\sigma_{n} \ldots \sigma_N} \equiv T^{(a_{n-1} \sigma_n)(\sigma_{n+1} \ldots \sigma_N)}
\end{equation}
where we assume that $\mathrm{dim}(a_0) = \mathrm{dim}(a_N) = 1$. Each time we do this, the SVD will give us a new tensor on the left whilst chipping away one physical index of the remaining tensor. The final result of this is
\begin{equation}
T^{\sigma_1 \sigma_2 \ldots \sigma_N } = A^{\sigma_1}_1 A^{\sigma_2}_2 \ldots A^{\sigma_N}_N = \sum_{ \mathbf{a}} (A_1^{\sigma_1})_{a_1} (A_2^{\sigma_1})_{a_1 a_2} \ldots (A_N^{\sigma_1})_{a_{N-1}}. \label{eq:mps}
\end{equation}
This is known as a MPS~\cite{tensor_networks} or a tensor train~\cite{doi:10.1137/090752286}. Each matrix is labelled by a bit $\sigma_i$ and its elements are $(A^{\sigma_i}_i)_{a_i a_{i+1}}$. The dimension of these matrices is given by $(\chi_{i-1}, \chi_i)$ where
\begin{equation}
\chi_i = \mathrm{dim}(a_i) = \min \{ 2^i, 2^{N-i} \}   \label{eq:untruncated_bond_dimension}
\end{equation}
where $\chi_i$ is known as the \textit{local bond dimension} of the $i$th bond. The left and right end tensors are vectors, as their respective left and right bonds have dimension one. The maximum bond dimension is located in the middle and is given by $2^{N/2}$.

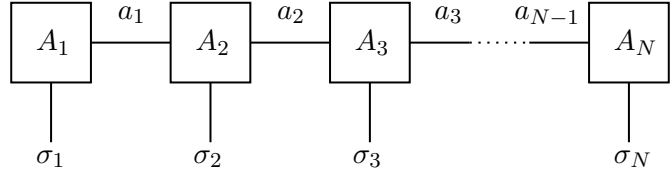
\begin{figure}[t]
\begin{center}
\tikzset{every picture/.style={line width=0.75pt}} 

\begin{tikzpicture}[x=0.75pt,y=0.75pt,yscale=-1,xscale=1]

\draw   (60,40) -- (100,40) -- (100,80) -- (60,80) -- cycle ;
\draw   (140,40) -- (180,40) -- (180,80) -- (140,80) -- cycle ;
\draw   (220,40) -- (260,40) -- (260,80) -- (220,80) -- cycle ;
\draw   (350,40) -- (390,40) -- (390,80) -- (350,80) -- cycle ;
\draw    (100,60) -- (140,60) ;
\draw    (180,60) -- (220,60) ;
\draw    (260,60) -- (290,60) ;
\draw    (320,60) -- (350,60) ;
\draw  [dash pattern={on 0.84pt off 2.51pt}]  (290,60) -- (320,60) ;
\draw    (80,80) -- (80,110) ;
\draw    (160,80) -- (160,110) ;
\draw    (240,80) -- (240,110) ;
\draw    (370,80) -- (370,110) ;

\draw (71,52.4) node [anchor=north west][inner sep=0.75pt]    {$A_{1}$};
\draw (151,52.4) node [anchor=north west][inner sep=0.75pt]    {$A_{2}$};
\draw (231,52.4) node [anchor=north west][inner sep=0.75pt]    {$A_{3}$};
\draw (361,52.4) node [anchor=north west][inner sep=0.75pt]    {$A_{N}$};
\draw (71,112.4) node [anchor=north west][inner sep=0.75pt]    {$\sigma _{1}$};
\draw (150,112.4) node [anchor=north west][inner sep=0.75pt]    {$\sigma _{2}$};
\draw (230,112.4) node [anchor=north west][inner sep=0.75pt]    {$\sigma _{3}$};
\draw (361,112.4) node [anchor=north west][inner sep=0.75pt]    {$\sigma _{N}$};
\draw (112,40.4) node [anchor=north west][inner sep=0.75pt]    {$a_{1}$};
\draw (192,40.4) node [anchor=north west][inner sep=0.75pt]    {$a_{2}$};
\draw (271,40.4) node [anchor=north west][inner sep=0.75pt]    {$a_{3}$};
\draw (311,40.4) node [anchor=north west][inner sep=0.75pt]    {$a_{N-1}$};

\end{tikzpicture}
\caption{A matrix product state is a special type of tensor network constructed from a set of tensors contracted into a one-dimensional chain. Here, each tensor core $(A_i)^{\sigma_i}_{a_{i-1} a_i}$ is represented as a square and each leg corresponds to an index. Two cores connected via a leg are contracted across the corresponding shared indices. \label{fig:mps}}
\end{center}
\end{figure}

We refer to the tensors $(A_i^{\sigma_i})_{a_i a_{i+1}}$ as \textit{tensor cores} and for a fixed $i$ and $\sigma_i$ they can be interpreted as matrices, hence the name MPS. Pictorially, the total MPS can be represented as a diagram shown in Fig.~\ref{fig:mps} where each square represents one of the tensor cores, whose three legs represent each of its three indices. The end tensors are viewed as vectors, as their outer bond dimension is one and we do not draw these. Any two cores joined by a leg represents a contraction of those two cores with respect to the shared bond index. To avoid messy notation, we drop the index labelling the lattice site and write our tensors as $A^{\sigma_i}_{a_i a_{i+1}}$ which is to be understood as the tensor located on site $i$. 

Note, approximations of the SVD, e.g., QR decomposition, could be utilised in-place of SVD and may be beneficial for compressing particularly large datasets. It is likely that an optimal compression strategy for large datasets could be to utilise one of SVD, QR decomposition, or other approximations at different stages of the process in order to minimise information loss while perfroming compression with suitable computational resources.

\subsection{Compression of matrix product states \label{sec:compression}}

The goal is to encode data into an MPS in a way that requires less memory than storing the raw data. When using MPSs, we do not store the entire tensor $T^{\sigma_1 \sigma_2 \ldots \sigma_N}$ in memory because this is equivalent to storing the raw data. Instead, we store the uncontracted tensor cores. If we wish to return the raw data, we can perform the contraction of Eq.~\eqref{eq:mps} to return the tensor. As each tensor core has dimension $\chi_{i-1} \times 2 \times \chi_i$ then storing each core requires storage of $2 \chi_{i-1} \chi_i$ numbers. The total amount of numbers to store is 
\begin{equation}
d = \sum_{i = 1}^N 2 \chi_{i-1} \chi_i, \label{eq:storage}
\end{equation}
where $\chi_0 = \chi_N = 1$. In this study, we compress our MPSs by truncating the singular values each time we perform an SVD, which reduces the local bond dimensions $\chi_i$ below their original size in  Eq.~\eqref{eq:untruncated_bond_dimension}, to give us an object that approximates the original tensor. There are two common approaches to truncation that we now outline.

\begin{figure}[t]
\begin{center}
\includegraphics[width = \textwidth]{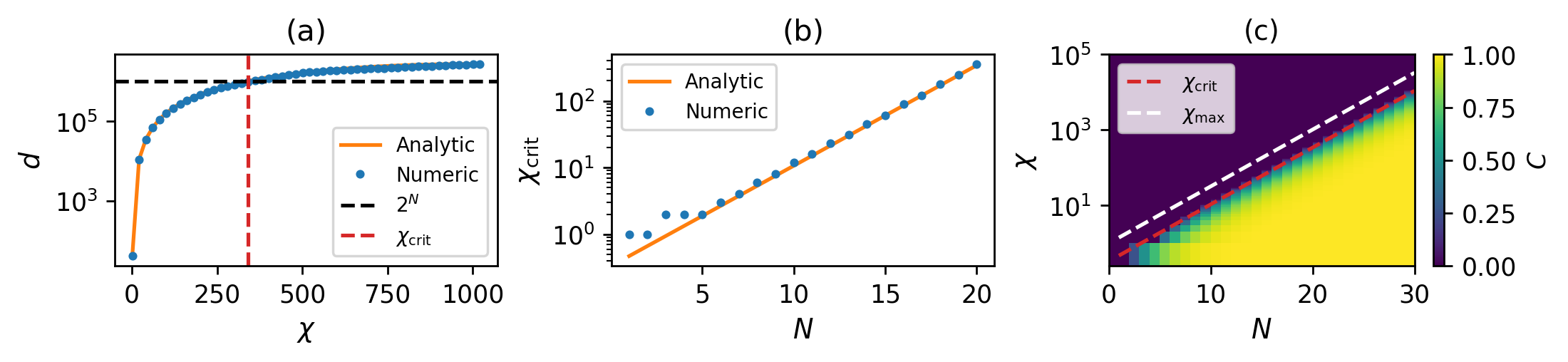}
\end{center}
\caption{In each subplot, we consider compression via the first method of Sec.~\ref{sec:compression} whereby we truncate the maximum bond dimension. (a) A comparison of the memory requirements for storing an MPS of size $N = 40$ vs. the maximum bond dimension. Here we plot the exact value obtained through numerics and compare it to the upper bound of Eq.~\eqref{eq:memory_storage}. The compression threshold is equal to $2^N$ and is the memory required to store the uncompressed function. The critical value is given by $\chi_\text{crit} = 2^{N/2}/3 $ and is the largest value of $\chi$ corresponding to the compression threshold (b) A comparison of the compression critical value $\chi_\text{crit}$ estimated from the analytical function vs. the true value. We see that it is an excellent approximation as $N$ gets large. (c) The compression achievable given an $N$-site MPS with a maximum bond dimension of $\chi$, where $C$ is defined in Eq.~\eqref{eq:compression}. The red line represents the critical threshold above which compression is not possible, whilst the white line corresponds to the maximum bond dimension for any tensor. \label{fig:compression}  }
\end{figure}

The first approach is to keep the $\chi$ largest singular values, often referred to as a maximum bond dimension. In which case the local bond dimension of Eq.~\eqref{eq:untruncated_bond_dimension} is modified to 
\begin{equation}
\chi_i = \min \{2^i, 2^{N-i} , \chi \} .\label{eq:truncated_bond_dimension}
\end{equation}
Note, if $\chi \geq 2^{N/2}$ then all singular values are retained and the MPS representation is exact. This method is a good choice if we wish to control the size of the data after compression. An example of the local bond dimensions when compressing this way is shown in Fig.~\ref{fig:compression}(c).  The local bond dimensions obey 
\begin{equation}
\chi_i = \min \{ 2^i, 2^{N-i} , \chi \} = \begin{cases}
2^i & 1 \leq \lfloor \log_2 \chi \rfloor \\
\chi &  \lfloor \log_2 \chi \rfloor < i < N  -  \lfloor \log_2 \chi \rfloor \\
2^{N - i} & \text{otherwise}
\end{cases}
\end{equation}
Plugging this into Eq.~\eqref{eq:storage} and assuming $\log_2(\chi) \in \mathbb{Z}$, then we get
\begin{equation}
d(\chi) = \frac{20}{3} \chi^2 + 2(N - 2 \log_2 \chi - 2) \chi^2 - \frac{8}{3} .
\label{eq:memory_storage}
\end{equation}
For $\log_2(\chi) \notin \mathbb{Z}$, this function provides an upper bound as shown in Fig.~\ref{fig:compression}(a). 

Compression is obtained if $d < 2^N$, where $2^N$ is the number of elements to store the original data in its uncompressed form. We can use Eq.~\eqref{eq:memory_storage} to find the critical bond dimension, defined as the largest value of $\chi$ for compression, however this must be solved numerically. We find the critical value is approximately
\begin{equation}
\chi_\text{crit} \approx \frac{2^{N/2}}{3}, \label{eq:chi_crit}
\end{equation}
below which we have compression. In Figs.~\ref{fig:compression}(a) and (b) we compare this against the numerics and see that it is a good approximation as $N$ scales. If we are below this threshold, we define the compression as
\begin{equation}
C = 1 - \frac{d}{2^N}, \label{eq:compression}
\end{equation}
where $d$ is given by Eq.~(\ref{eq:storage}). This quantity is sometimes known as the space saving. In Fig.~\ref{fig:compression}(c) we show the compression achievable for various $N$ and maximum bond dimension $\chi$ as we generate random data sets of size $2^N$. 

The second approach to compress an MPS is to retain singular values above a minimum threshold $s_\text{min}$. In this case, the local bond dimensions will not follow Eq.~\eqref{eq:truncated_bond_dimension} and can vary across the MPS as this method depends upon the details of the data stored in the tensor, as shown in Fig.~\ref{fig:compression_time}(c). One way we can do this by taking the threshold to be a fraction of the largest singular value as $s_\text{min} = k s_\text{max}$ for $k \in [0,1]$. This method is a good choice if we wish to control the error of a compression, and want to ensure only relevant singular values are retained throughout the MPS.  Unlike the first method, there is no simple formula for $d$ so this must be calculated numerically and different data compressed to the same $k$ will have different sizes.

\subsection{Compression complexity}
\begin{figure}
\begin{center}
\includegraphics[width = \textwidth]{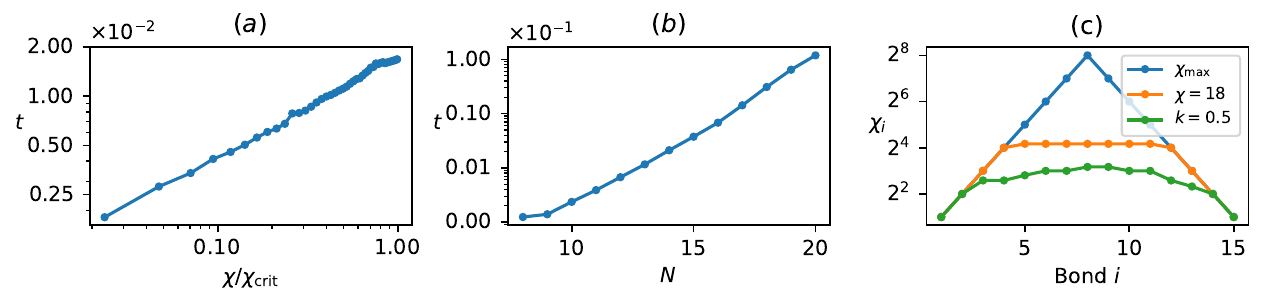}
\end{center}
\caption{(a) Compression time of random data sampled in the range $[0,1]$ of length $2^N$ for $N = 16$ versus the maximum bond dimension $\chi$, where $\chi_\text{crit} = 2^{N/2}/3 = 85$. This displays polynomial scaling as this is a log-log plot. (b) Compression time scaling for random data sampled in the range $[0,1]$ of length $2^N$ for fixed $\chi = 40$ for various $N$. This displays exponential scaling. (c) The local bond dimensions $\chi_i$ for an MPS encoding a random Fourier series with 20 modes of length for $N = 16$ before and after truncating the maximum bond dimension to $\chi = 18$ or when we compress to a singular value threshold of $k s_\text{max}$ for $k = 0.5$, where $s_\text{max}$ is the maximum singular value of a given SVD. \label{fig:compression_time}} 
\end{figure}

In this section we derive the time scaling of the compression algorithm. Given a matrix of size $m \times n$, the SVD has complexity $O( \min \{ m^2 n,n^2 m \} )$. When decomposing an $N$th degree tensor into a MPS, we perform $N-1$ SVDs. For the $i$th SVD, the $m \times n$ matrix has the dimension $m =2 \chi_{i-1}$ and $ n = 2^{N-i}$ where the bond dimensions are given by Eq.~\eqref{eq:truncated_bond_dimension}. As in this case we have
\begin{equation}
\min \{ 2 \chi_{i-1}, 2^{N-i} \} = \begin{cases}
2 \chi_{i-1} & 1 \leq i < N - \lfloor \log_2 \chi \rfloor \\
2^{N-i} & \text{otherwise}
\end{cases},
\end{equation}
it follows that the complexity of the $i$th SVD is
\begin{equation}
C_i   \leq  \begin{cases}
4 \chi^2 2^{N-i} & 1 \leq i < N - \lfloor \log_2 \chi \rfloor \\
2 \chi 4^{N-i} & \text{otherwise}
\end{cases},
\end{equation}
where we used the fact the local bond dimension obeys $\chi_i \leq \chi$. The total complexity of all SVDs for an $N$th degree tensor is given by the sum of these terms, then we get
\begin{equation}
\sum_{i = 1}^{N-1} C_i  \leq  \frac{8}{3} \left( 3 \chi^3 + \chi^2 -  \chi \right) +  2^{N+2} \chi^2   = O(\chi^3) + O(2^N \chi^2),
\end{equation}
where for simplicity we have assumed $\log_2(\chi) \in \mathbb{Z}$. In Fig.~\ref{fig:compression_time} we show this scaling behaviour for random data of length versus the bond dimension and the size of the lattice for data sampled from randomly $[0,1]$, and we additionally show how this affects the local dimension across the MPS after compression.

\section{Simple test: Random Fourier series}
\subsection{Exact MPS encodings \label{sec:1d-fourier}}
Some functions have exact MPS expressions, allowing for an efficient encoding in MPS form without having to use SVDs. For example, consider the complex exponential $f(x) = e^{i k x}$. If we insert in the binary representation of $x$ of Eq.~(\refeq{eq:binary_to_decimal}) then we have
\begin{equation}
e^{ik x(\sigma)} = \prod_{n = 1}^N e^{ik \frac{\sigma_n}{2^n} }.
\end{equation} 
We can read off the MPS form of this function as
\begin{equation}
T^{\sigma_1 \sigma_2 \ldots \sigma_N} = \prod_{n = 1}^N A^{\sigma_n}, \quad A^{\sigma_n} = e^{ik \frac{\sigma_n}{2^n}}. \label{eq:complex_exponential}
\end{equation}
This is an MPS of bond dimension of $1$ as each tensor core is just a number. Initially, to store this function on the discretised grid we needed to store $2^N$ numbers. After, we need to store $N$ tensors, each of which contains $2$ elements. Therefore, storing the entire MPS requires storage of $2N$ numbers. This is an exponential decrease in memory requirements and is independent of the frequency $k$.

Now suppose we have a periodic function $f(x)$ on the range $[0,1]$. This can be written as a Fourier series as
\begin{equation}
f(x) = \sum_{k \in \Omega} c_k e^{i k x}, \label{eq:fourier_series}
\end{equation}
where $\Omega$ is a set of frequencies of the form $k = 2 n \pi$ for $n \in \mathbb{Z}$ and we denote $M = |\Omega|$ as the number of complex Fourier modes. We can write the MPS representation of this by first replacing the functions on either side of the equality by their tensor representations as
\begin{equation}
T^{\sigma_1 \sigma_2 \ldots \sigma_N}  = \sum_{k \in \Omega} c_k T_k^{\sigma_1 \sigma_2 \ldots \sigma_N},
\end{equation}
where $T$ and $T_k$ are the tensor representations of $f(x)$ and $e^{ikx}$ respectively. If we replace each tensor with its MPS form, we have
\begin{equation}
A^{\sigma_1} A^{\sigma_2} \ldots A^{\sigma_N} = \sum_{k \in \Omega} c_k A^{\sigma_1}_k A^{\sigma_2}_k \ldots A^{\sigma_N}_k,
\end{equation}
where $A_k^{\sigma_i}$ are given by Eq.~(\ref{eq:complex_exponential}). It can be shown that tensor cores of a sum of MPSs are constructed from the tensor cores of each MPS in the sum as
\begin{subequations}\label{eq:fourier_tensor_cores}
\begin{align}
A^{\sigma_1} & = \begin{pmatrix} c_{k_1} A^{\sigma_0}_{k_1} & c_{k_2} A^{\sigma_0}_{k_2} & \ldots \end{pmatrix}^T , \label{eq:tensor_cores_1}\\
A^{\sigma_n} &  =  \begin{pmatrix}
A^{\sigma_n}_{k_1} & &   \\
& A^{\sigma_n}_{k_2} &    \\
& & \ddots    \\
\end{pmatrix}, \quad 1 < n < N, \label{eq:tensor_cores_2} \\
A^{\sigma_N} & = \begin{pmatrix} A^{\sigma_N}_{k_1} & A^{\sigma_N}_{k_2} & \ldots  \end{pmatrix}. \label{eq:tensor_cores_3}
\end{align}
\end{subequations}
where we have used matrix notation to represent the components of the tensor cores $A^{\sigma_i}_{a_{i-1} a_i}$ with respect to the bond indices $a_{i-1}$ and $a_i$~\cite{tensor_networks}. The maximum bond dimension of this exact encoding is $\chi = M$ and is independent of the number of sites $2^N$. This is an upper bound however: if we add two MPSs with bond dimensions $\chi_1$ and $\chi_2$, then the bond dimension of the sum is bounded as $ \leq \chi_1 + \chi_2$. An example that does not saturate this bound is when we add an MPS to itself, then its bond dimension should not double as adding it to itself is equivalent to scaling by a factor of two, which does not increase bond dimension~\cite{tensor_networks}.

\begin{figure}
    \centering
    \includegraphics[width=\linewidth]{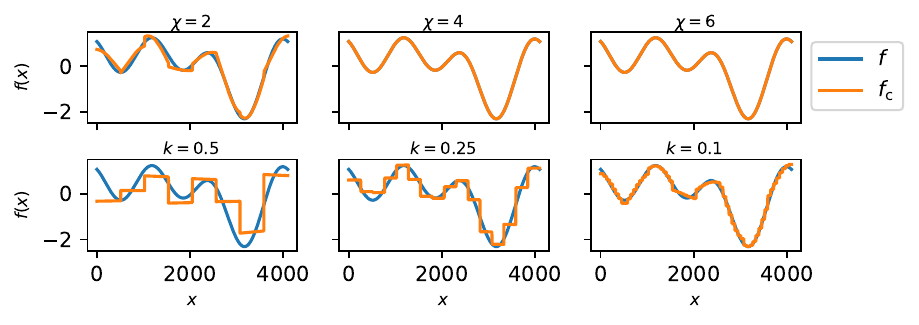}
    \caption{(Top row) A comparison of a function $f$ with six Fourier modes for $N = 12$ to the MPS-compressed version $f_\text{c}$ for various maximum bond dimensions $\chi$. We obtain $f_\text{c}$ by fully contracting the MPS back into a vector, as in Eq.~\eqref{eq:mps}. Lossless encoding is when $\chi = 6$. (Bottom row) A comparison where we take the singular value threshold as $k s_\text{max}$, where $s_\text{max}$ is the maximum singular value of a given SVD. }
    \label{fig:example_fourier_compression}
\end{figure}

In practice we will not know what the Fourier coefficients are, so we cannot construct the MPS of our data using this exact form. In fact, if we knew the Fourier modes we would not bother constructing the MPS this way as we could simply store the Fourier spectrum. Instead, we are presented with a data set and perform the numerical steps outlined in Secs.~\ref{sec:background} and \ref{sec:compression} by reshaping the data into a tensor, applying successive SVDs, and truncating the singular values to a bond dimension of $\chi$ or a given threshold. The analytic form of Eq.~\eqref{eq:fourier_tensor_cores} tells us that an exact MPS encoding can be obtained if the bond dimension $\chi$ is equal to the number of Fourier modes and this result still applies if we compress using the numerical method. Therefore, an exact encoding of a Fourier series with $M$ modes into an MPS is guaranteed if we compress by truncating the bond dimension to
\begin{equation}
\chi = \min \{M, 2^{N/2} \}.
\end{equation}
In Fig.~\ref{fig:example_fourier_compression} we show an example of a function with six Fourier modes being compressed via the two methods. We see that even if $\chi$ does not satisfy this requirement, so the compression is lossy, the approximation is good. If we combine this with the critical value for compression from Eq.~\eqref{eq:chi_crit}, then \textit{lossless compression} of a Fourier series containing $M$ Fourier modes is guaranteed if
\begin{equation}
M  \leq \chi \lesssim  \frac{2^{N/2}}{3}. \label{eq:fourier_compression_threshold}
\end{equation}
In practice, this constraint can be relaxed. If the number of modes $M$ is large, then $\chi$ can be smaller than $M$ which is seen in Fig.~\ref{fig:error_scaling}(a), however this value of $\chi$ may lie in the region of no compression.

The Nyquist limit tells us that if we have a periodic signal with frequency $f$ then we must sample it with a frequency of $f_s = 2f$ in order to resolve its frequency. As discretising a continuous function on a lattice with lattice spacing $\Delta x$ is equivalent to sampling the function with a sampling frequency of $f_s = 1/\Delta x$, if we discretise a function into $2^N$ points, then the frequency range we can resolve is given by
\begin{equation}
f < 2^{N-1} .
\end{equation} 
Any frequencies $\geq 2^{N-1}$ cannot be resolved. A maximum frequency of $\Lambda = 2^{N-1} - 1$ means the most general Fourier series has the frequencies $\Omega = \{ -\Lambda ,\ldots , \Lambda \}$, in which case $M = 2\Lambda + 1 = 2^{N} - 1$. This is clearly above the compression threshold of Eq.~\eqref{eq:fourier_compression_threshold}, so we conclude that not every possible Fourier series can be losslessly compressed in MPS form, as some functions may contain all frequencies up to the cutoff. In practice, we apply Orszag's $2/3$ rule by filtering out the top third of frequencies due to the aliasing error that is introduced which will bring down the threshold.

Note that this does \textit{not} mean that we cannot losslessly encode a function with a high frequency, as the compression condition is concerned with the number of frequencies and not what they are; we can still compress a high frequency signal up to the Nyquist limit with no loss, as long as the total number of frequencies is not too high. It also does not mean that if a function has a high number of Fourier modes it cannot be compressed losslessly. Polynomials have a large number of Fourier modes, yet there exists an exact encoding of these functions into MPS of low bond dimension: given a univariate polynimial of degree $p$ we can encode it in an MPS of bond dimension of at most $p+1$~\cite{ali2023approximation} which, similarly to encoding Fourier series, exploits the structure of the function to find an efficient encoding. In CFD, it is unlikely our data will take this form however, so we resort to random Fourier series and the bounds of Eq.~\eqref{eq:fourier_compression_threshold}.

\subsection{Compression results}

We now investigate the MPS compression approach via compression of a random Fourier series. We encode our Fourier series of length $2^N$ into an MPS of length $N$ using the two methods outlined in Sec.~\ref{sec:compression} and reconstruct the function by contracting its corresponding MPS back into a vector using Eq.~\eqref{eq:mps}. We define the error as
\begin{equation}
\epsilon = \frac{\parallel f - f_\text{c} \parallel}{\parallel f \parallel},
\end{equation}
where $\parallel \cdot \parallel $ is the $L_2$ norm, $f_\text{c}$ is the compressed function, and $f$ is the original uncompressed function. We generate random real Fourier series of the form of Eq.~(\ref{eq:fourier_series}) by choosing a random set of distinct frequencies $\Omega$ below the Nyquist limit of the given size and sampling the real and imaginary values of the Fourier coefficients each the range $[0,1]$. 

\begin{figure}
\begin{center}
\includegraphics[width=0.8\textwidth]{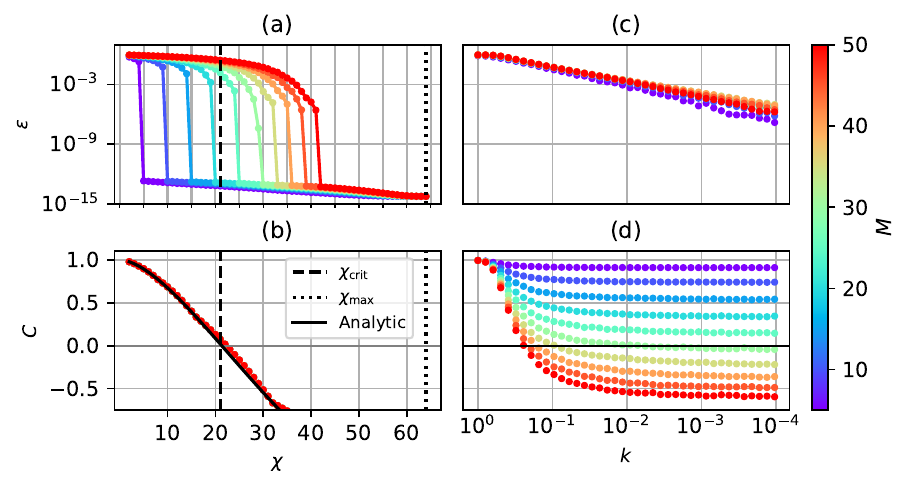}
\caption{We apply the two methods of compressing data into an MPS, either truncating to a maximum bond dimension of $\chi$ or a retaining singular values above a threshold $k s_\text{max}$ for $ k \in [0,1]$ outlined in Sec.~\ref{sec:compression}, to random Fourier series containing $M$ modes on a grid of size $2^N$ for $N = 12$. Each coloured line corresponds to a given number of Fourier modes $M \in \{5,10,15,\ldots,50\}$ and is repeated 500 times. (a) The average error of the MPS compression as a function of the maximum bond dimension. For small $M$ we see the expected behaviour that the encoding is lossless when $\chi = M$. For large $M$ this bound becomes looser and a smaller bond dimension is sufficient for a lossless encoding, however this may lie beyond the threshold for compression $\chi_\text{crit} \approx 2^{N/2}/3$. The maximum bond dimension is given by $\chi_\text{max} = 2^{N/2}$. (b) The compression of the MPS as a function of the maximum bond dimension. We see that as the compression is a property of the MPS and not the data that is encoded in it, the compression is identical for all cases. The analytic line is $1 - d(\chi)/2^N$ where $d(\chi)$ is from Eq.~\eqref{eq:memory_storage}. (c) The average error of the MPS as a function of the singular value threshold. We see that the error decreases exponentially with $k$ and is insensitive to the number of modes. (d) The average compression of the MPS as a function of the singular value threshold. \label{fig:error_scaling}}
\end{center}
\end{figure}

In Fig.~\ref{fig:error_scaling} (a) we compare the error versus the maximum bond dimension when we compress via the first method. First, we see that for a small number of Fourier modes, if the bond dimension obeys $\chi < M$ then we have lossy compression and error is introduced, however the error decreases with increasing $\chi$. Once $\chi \geq M$ we see the error drops sharply, signifying a lossless compression as expected from the analytics of Sec.~\ref{sec:1d-fourier}.  We also see that as the number of Fourier modes gets large, then the lower bound of Eq.~\eqref{eq:fourier_compression_threshold} for lossless encoding can be relaxed, as a  bond dimension smaller than $M$ can be sufficient for an exact encoding. However, this may lie beyond the compression threshold $\chi_\text{crit} = \frac{2^{N/2}}{3}$ and would therefore not correspond to compression. In Fig.~\ref{fig:error_scaling}(b) we compare the compression versus the maximum bond dimension via the first method. We see the compression is independent of the number of Fourier modes, as compression here is independent of the data and instead of a property of the dimension of the tensor cores. Note that if the bond dimension is equal to the maximum bond dimension of $\chi_\text{max} = 2^{N/2}$ then we can encode any data exactly, but it will not be compressed.

In Fig.~\ref{fig:error_scaling} (c) we compare the error versus the singular value threshold when we compress via the second method. As our data is not normalised, the singular values are not bounded between $[0,1]$ so we introduce a threshold given by a fraction of the maximum singular value of each SVD, in other words we retain the singular value $s$ if $s \geq k s_\text{max}$ where $k \in [0,1]$ and $s_\text{max}$ is the maximum singular value. We see that in this case, the error is largely insensitive to the number of Fourier modes present in the data and decays exponentially as $k$ decreases. On the other hand, in Fig.~\ref{fig:error_scaling}(d) we see the compression achievable is now dependent on the number of Fourier modes, and we see that as the number of modes increases, the compressibility of the data decreases using this method and there is no sharp cutoff.

Fig.~\ref{fig:error_scaling} highlights the differences between the two methods of compression: truncation of the maximum bond dimension fixes the compression, whilst truncating the singular values approximately fixes the error. The choice of method is dictated by whether we value memory or precision in our compression. Moreover, a hybrid compression where we truncate singular values as well as introducing a maximum number of singular values can also be employed. This includes enforcing the different truncation approaches simultaneously on different parts of the MPS, which will be explored in future work.

\section{Application to fluid dynamics}\label{sec:application-1d-burgers}
\begin{figure}
\begin{center}
\includegraphics[width=0.98\textwidth]{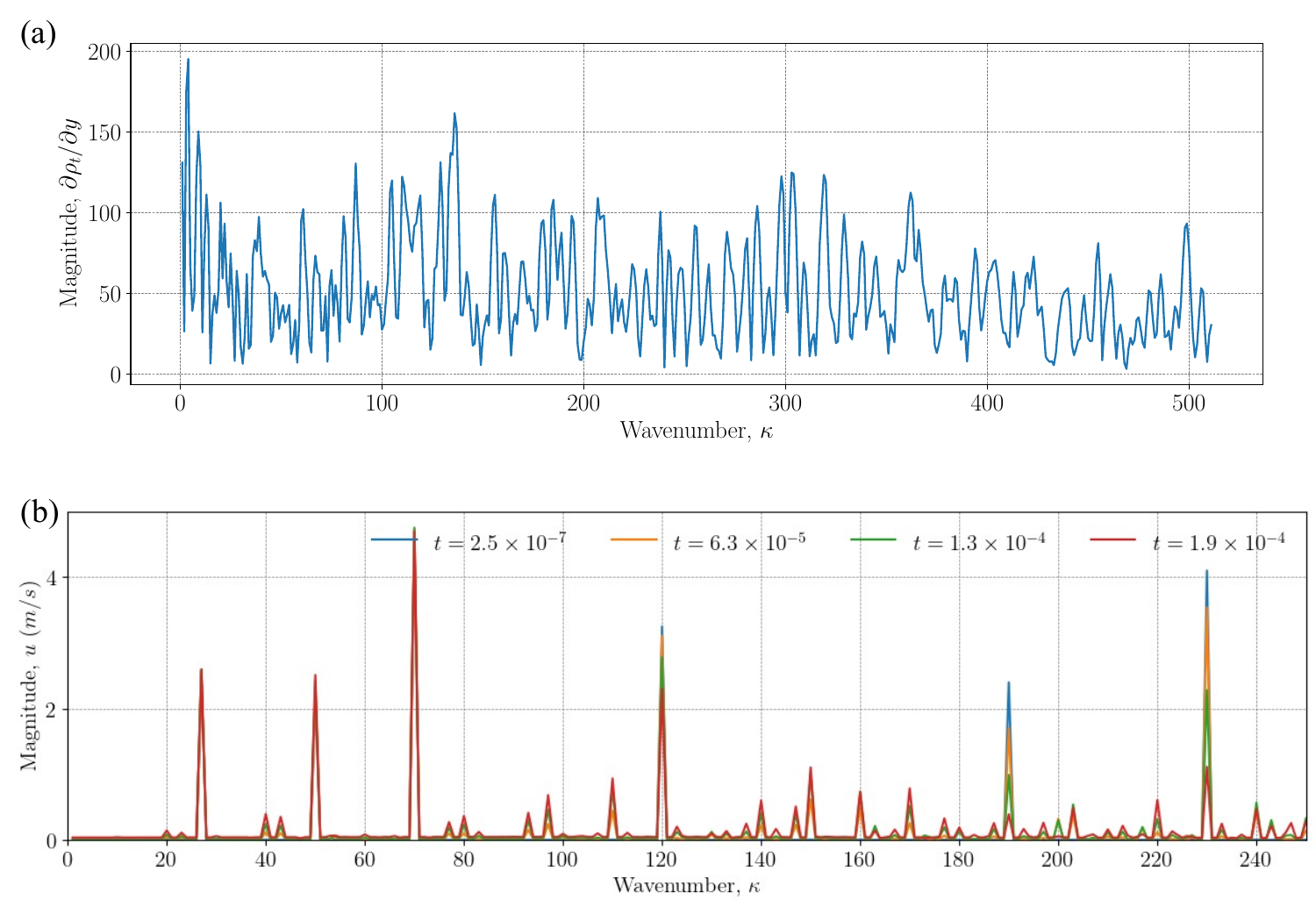}
\end{center}
\caption{(a) Spectra of the representative 1D data shown in figure~\ref{fig:flow_non-iso-sst}~(b). (b) Spectra of the 1D problem used in the current work \cite{gopalakrishnan2026tensor}. The flow is initialized using the form in \cite{walters2018favre}.
\label{fig:flow_1D-spectra}}
\end{figure}

The fluids data depicted in Fig.~\ref{fig:flow_non-iso-sst} is broad-banded---depicted by the spectra shown in figure~\ref{fig:flow_1D-spectra}~(a)---and the preceding analyses show that the tensor-network approach to compression is attractive for broad-banded data.  Another fundamental feature of fluid flows, though, is that they evolve in time.  Common fluid flows are described by the Navier--Stokes equations, which are coupled non-linear partial differential equations in space and time, and which are inherently multi-dimensional because of the effect of pressure.  A simpler and commonly used model problem for fluid flows is Burgers' equation \cite{burgers1940}
\begin{equation}
\frac{\partial u}{\partial t} + u \frac{\partial u}{\partial x} = \nu \frac{\partial^2 u}{\partial x^2},
\label{eq:burgers}
\end{equation}
where $u = u(x,t)$ is a one-dimension velocity field, $\nu$ is the kinematic viscosity, $t$ is time, and $x$ is the spatial coordinate.
Without the pressure gradient intrinsic in Navier--Stokes, Burgers allows the study of the development of an ever-broader range of frequencies in time that must be dealt with by any compression algorithm.  
For this research, we chose a representative 1D flow that has similar broad-band features to the SST flow (figure~\ref{fig:flow_non-iso-sst}) as shown in figure~\ref{fig:flow_1D-spectra}~(b), but not as complex. This flow is initialized based on the formulation in Ref.~\cite{walters2018favre} and the dynamics solved based on the Burgers' equation. A preliminary tensor-network-based modeling of this use case was recently performed \cite{gopalakrishnan2026tensor}.

The preceding section related the number of Fourier modes to an upper bound on the maximum bond dimension $\chi$ required for lossless compression. In order to gauge the applicability of this to computational fluid dynamics, we need to relate this to the Reynolds number. Kolmogorov scaling tells us we require $ \sim \mathrm{Re}^{3/4}$ grid points given a Renynolds number $\mathrm{Re}$~\cite{10.1098/rspa.1991.0075}. This, together with the Nyquist limit, tells us that the number of Fourier modes goes as $M \sim \mathrm{Re}^{3/4}$. If we combine this with the bounds of Eq.~\eqref{eq:fourier_compression_threshold} we can expect lossless compression when truncating the bond dimension as long as
\begin{equation}
\mathrm{Re}^{3/4} \leq \chi \leq \chi_\text{crit}.
\end{equation}
If we choose to compress with a singular value threshold instead, then the dependence on $k$ is difficult to obtain, instead given a $k$ and an implementation we can obtain the maximum bond dimension numerically and revert back to the relation to $\chi$ for insight.

\subsection{Correlation of length scales}
\begin{figure}
\begin{center}
\includegraphics[width=0.8\textwidth]{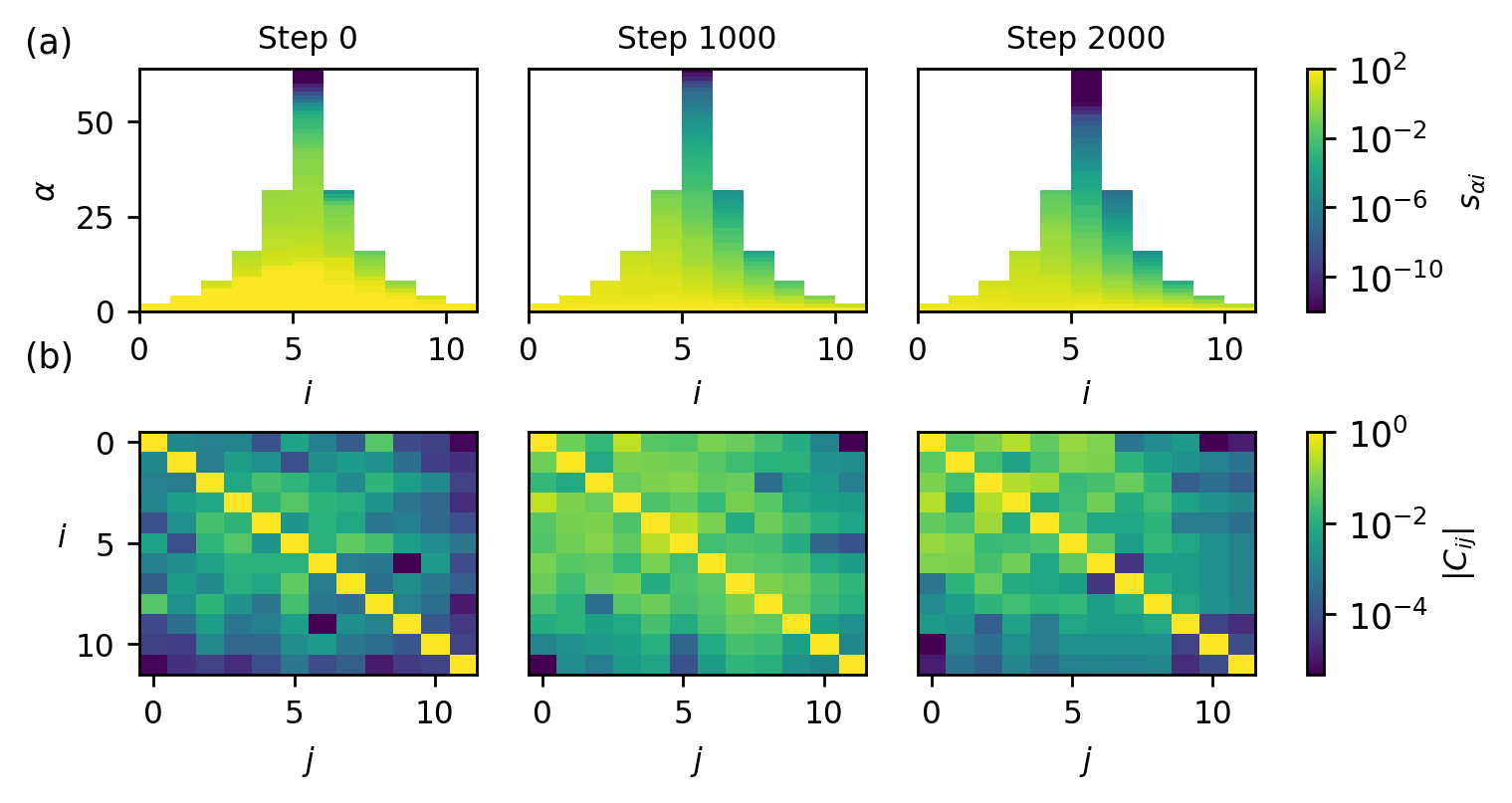}
\caption{(a) The singular value distribution along the MPS for DNS solutions to the Burger's equation for $N = 14$ for various times with a time step size of $\Delta t = 10^{-8}$ and viscosity $\nu = 10^{-6}$ initialised in the input state of Ref.~\cite{walters2018favre} for various time steps. For each bond index $i$, we list the distribution of singular values vertically indexed by $\alpha$, where $s_{\alpha i }$ is defined as the $\alpha$th singular value of the $i$th bond. We see as the system evolves in time the amount of non-trivial singular values remains relatively low. (b) The correlation matrices $C_{ij} = \langle Z_i Z_j \rangle - \langle Z_i \rangle \langle Z_j \rangle$. The correlation between length scales remains small over time, suggesting that the structure of an MPS is well-suited to this data.
\label{fig:correlation_matrices}}
\end{center}
\end{figure}
Eq.~\eqref{eq:burgers} can be solved on the range $[0,1]$ by discretising it into $2^N$ points, replacing spatial derivatives with finite differences and performing a finite time stepping such as Euler or Runge-Kutta. Euler stepping here is given by
\begin{equation}
\begin{aligned}
u(x, t + \Delta t) & := u(x,t) + \Delta t \frac{\partial u(x,t)}{\partial t} \\
& = u(x,t) + \Delta t \left(  \nu \frac{\partial^2 u}{\partial x^2} - u \frac{\partial u}{\partial x} \right). \label{eq:euler_step}
\end{aligned}
\end{equation}
Due to the energy cascade in fluid dynamics, energy is transferred between similar length scales, so widely-separately length scales are not expected to interact directly~\cite{10.1098/rspa.1991.0075}. As each tensor core of a matrix product state corresponds to a different length scale due to the binary encoding, see Eq.~\eqref{eq:binary_to_decimal}, and the bond dimension between the cores carries the correlation, we expect that a low bond dimension MPS should be sufficient at representing fluid data.

In order to investigate this, we first encode the discretised velocity field vector at a given time $t$ obtained via Euler stepping into a tensor $T^{\sigma_1 \sigma_2 \ldots \sigma_N}$ with no compression, then we calculate the correlation matrix
\begin{equation}
C_{ij} = \langle   Z_i Z_j \rangle - \langle Z_i \rangle \langle Z_j \rangle
\end{equation}
where the expectation value is given by the contraction
\begin{equation}
    \langle Z_i \rangle = \sum_{\boldsymbol{\sigma}, \boldsymbol{\sigma}'} T^{\sigma_1 \sigma_2 \ldots \sigma_i'\ldots  \sigma_N} Z^{\sigma_i' \sigma_i} T^{\sigma_1 \sigma_2 \ldots \sigma_i \ldots \sigma_N}, \quad Z= \begin{pmatrix} 1 & 0 \\ 0 & -1 \end{pmatrix}
\end{equation}
and similarly for $\langle Z_i Z_j\rangle$. This quantity calculates the correlation between bits $\sigma_i$ and $\sigma_j$, and hence the corresponding length scales, and is inspired from techniques used in quantum computing to calculate the correlation between two qubits. Note that in order to do this we must normalise the velocity field vectors such that they have unit $L_2$ norm. This is to ensure the velocity field can be interpreted as a probability amplitude function allowing us to construct the correlation matrix. In Fig.~\ref{fig:correlation_matrices} we see the correlation matrix for solving the Burgers' equation for $N = 14$. We see that that there is weak correlation between differing length scales, giving strong evidence that an MPS is a suitable candidate for compression of this data.

\subsection{Compression of dynamical data}

If we initialise the field as a Fourier series with a discrete set of frequencies and evolve it in time, we see that additional Fourier modes are excited throughout the evolution. From the above analysis, this suggests that as the number of modes grows the MPS must increase in bond dimension in order to compress faithfully. In this section, we investigate further.

Let us write the velocity $u(x,t)$ on the range $[0,L]$ with the definition in terms of its Fourier series as
\begin{equation}
u(x,t) = \sum_{k \in \Omega} c_k e^{ikx}, 
\end{equation}
where $\Omega$ are the frequencies. If we substitute this into the Burgers' equation we arrive at the equation of motion for the Fourier modes
\begin{equation}
\frac{\partial c_k(t)}{\partial t} + \nu k^2 c_k(t) + \frac{ik}{2} f_k(t) = 0,
\end{equation}
where $f_k(t)$ arises from the non-linear term and is given by the discrete convolution
\begin{equation}
f_k(t) = \sum_q c_{k-q} c_q.
\end{equation}
This term couples Fourier modes together, which explains why new Fourier modes are excited as the Burgers' equation evolves. 

If at some given time $t$ the system has the discrete Fourier spectrum of frequencies $\Omega$, then the specturm takes the form $
c_k(t) = \sum_{q \in \Omega} a_q \delta_{kq}$. If we substitute this into $f_k(t)$ we get
\begin{equation}
f_k(t)  = \sum_{p,l \in \Omega} a_p a_l \delta_{k, p + l}, 
\end{equation}
which is a another discrete frequency spectrum with support on the set $ \Omega + \Omega$, where set addition is defined as $A + B = \{ a + b : a \in A, \ b \in B \}$. Note that the size of this set can be less than $2 M$ as some terms may be repeated. 

From the equation of motion of Eq.~(\ref{eq:time_step}), the frequency spectrum after a time $\Delta t$ is approximately
\begin{equation}
c_k( t + \Delta t ) \approx (1 -  \nu k^2 \Delta t) c_k(t)  - \frac{i k \Delta t}{2} f_k(t). \label{eq:time_step}
\end{equation}
If we were doing finite Euler stepping then this would be the frequency spectrum precisely, as by definition Euler stepping only ever includes the $O(\Delta t)$ term. This time stepping rule consists of adding together two spectra (up to some overall constants). These are the original spectra $c_k$ and the convolution spectra $k f_k$. We can use this to calculate how the number of Fourier modes increases with each Euler step. The set of frequencies evolves for each time step as
\begin{equation}
\Omega_{n+1} = \Omega_n\cup [( \Omega_n + \Omega_n ) \setminus \{ 0 \}].  \label{eq:update_rule}
\end{equation} 
The first term in the union is due to the linear diffusive term of the Burgers' equation which does not change the set of frequencies, whereas the second term is from addition of the non-linear convolution which is responsible for the additional frequencies. We subtract off the contribution of the frequency $ k = 0$ in the convolution as $f_0$ does not contribute due to the factor of $k$ multiplying it. For example, if we Euler step the initial state $u(x,0) = \sin(2 \pi x)$ then we find $|\Omega_n| = 2^{n+1}$. In addition, if we were to use a different time stepping scheme such as Runge-Kutta then the sequence would be different.

Whilst this rule is formally how the spectrum behaves with Euler stepping, in practice the spectrum will be seen to behave differently. The main reasons are that many of the Fourier modes are vanishingly small so will not be important, and when solving this numerically we will see aliasing in the spectrum due to the finite lattice effects that excites many other modes. Regardless, the key takeaway is that the complexity of the flow will change over time and, using the results of Fig.~\ref{fig:error_scaling}, the bond dimensions of MPS will need to adapt over time if we are to retain a level of precision in our compression. One way we can attempt this is to compress our data using the singular value threshold discussed in Sec.~\ref{sec:compression} instead of fixing the bond dimension. In this case, the bond dimension and compression is dynamic while the error remains approximately constant. If we fix the maximum bond dimension, the error will be dynamic instead whilst the compression is fixed. 

\begin{figure}
\begin{center}
\includegraphics[width=\textwidth]{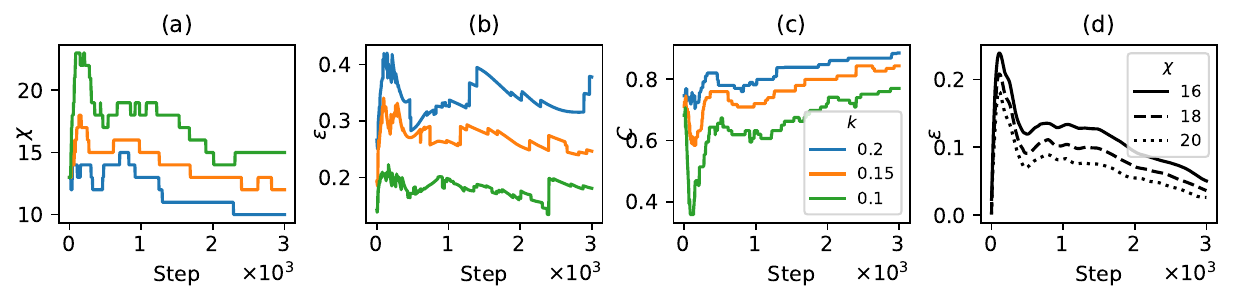}
\caption{The MPS properties over time for solutions to Burgers equation with initial condition with spectra as shown in Fig. with a system of $N= 12$, time step of $\Delta t = 10^{-6}$ and viscosity $\nu = 10^{-3}$. For each time step, we evolve using a DNS Euler stepping and compress the resultant solution into an MPS. (a)-(c) The maximum bond dimensions $\chi$, the compression error $\epsilon$ and the compression $C$ for various minimum singular value thresholds $s_\text{min} = k s_\text{max}$. As time evolves, the complexity of the solution changes which is reflected in a changing bond dimension and error. (d) We fix the bond dimensions to various values $\chi$. In this case, the bond dimensions to $\chi \in \{16,18,20\}$ and compression are fixed over time, with $C= 0.33$, $0.24$ and $0.13$ respectively. In this case the error is not constant and changes over time, but the general rule is that the larger $\chi$ the smaller the error. \label{fig:dynamics}}
\end{center}
\end{figure}

In Fig.~\ref{fig:dynamics} we compare the two methods of compression, by introducing a singular value threshold or introducing a maximum bond dimension, for DNS similation of the Burgers' equation for an initial condition consisting of 22 Fourier modes on a system with $N = 12$. For each time step, we compress the DNS solution into an MPS and analyse the compression. We see that after short times, the bond dimension grows with the former method, or the error grows with the latter method. This is because the number of Fourier modes grows quickly for initial times and hence requires larger MPS to compress. After a long time, the solution dissipates, which is reflected in the decrease in bond dimension and error of each method respectively. We see that we can achieve compression with low errors for thousands of time steps.

\subsection{Application to spectral methods}
One of powerful features of tensor networks is the ability to not only compress the data, but to process it efficiently once it is compressed. In quantum many-body physics for example, local information about the state of the system encoded into an MPS can be extracted without the need to contract it back to an exponentially large vector by contracting the MPS with a matrix product operator (MPO) representing some observable of interest~\cite{tensor_networks}. An MPO is similar to a MPS, except each tensor core has two or more binary indices instead of one, which represents a linear operator on the space of MPS. In fluid dynamics, it has been shown that the Navier-Stokes equations can be solved directly in tensor form, and operations such as finite differences and Hadamard products are translated into contraction with a suitable MPO~\cite{gourianov2022quantum, gourianov2025tensor, holscher2025quantum, lubasch2018multigrid}. This allows us to avoid contracting the MPS back into a vector to process it.

On the other hand, spectral methods can be employed in CFD instead~\cite{canuto2007spectral}. It has been shown that one can perform a quantum Fourier transform with tensor networks and can have orders of magnitude speedup over the fast Fourier transform~\cite{PRXQuantum.4.040318}, allowing us to analyse spectral data in CFD. However, some simple operations in real space, such as products, become more complicated in Fourier space and we need a way to perform them in tensor networks. As shown in Eq.~\refeq{eq:time_step}, the Burgers' equation after a Fourier transform contains a convolution of the form
\begin{equation}
f_k = \sum_{q \in \Omega} c_{k-q} c_q. \label{eq:convolution}
\end{equation}
Numerically, if we solve the problem on a discrete lattice of $L = 2^N$ sites with periodic boundary conditions, then Fourier space reduces to the Brillouin zone containing $L$ discrete frequencies and our spectral functions have the periodicity $c_{k + 2 \pi/a} = c_k$, where $a$ is the lattice spacing in real space, so the convolution is a \textit{periodic} convolution. The periodic convolution is a highly non-local operation in Fourier space as all modes are coupled together, so it is one of the bottlenecks for spectral methods as it scales as $O(L^2)$ for a grid of size $L$, which can be exponentially large. In this section, we demonstrate how one can perform a convolution by contracting with an MPO with a time scaling of $O(\log L)$ instead.

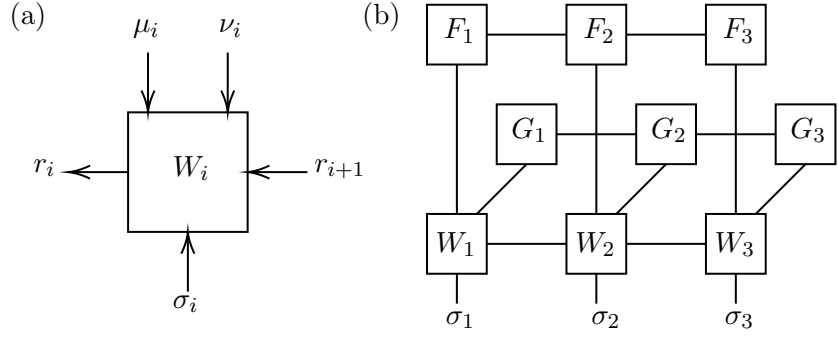
\begin{figure}
\begin{center}
\tikzset{every picture/.style={line width=0.75pt}} 

\begin{tikzpicture}[x=0.75pt,y=0.75pt,yscale=-1,xscale=1]

\draw   (65,64) -- (125,64) -- (125,124) -- (65,124) -- cycle ;
\draw    (95,154) -- (95,126) ;
\draw [shift={(95,124)}, rotate = 90] [color={rgb, 255:red, 0; green, 0; blue, 0 }  ][line width=0.75]    (10.93,-3.29) .. controls (6.95,-1.4) and (3.31,-0.3) .. (0,0) .. controls (3.31,0.3) and (6.95,1.4) .. (10.93,3.29)   ;
\draw    (75,34) -- (75,62) ;
\draw [shift={(75,64)}, rotate = 270] [color={rgb, 255:red, 0; green, 0; blue, 0 }  ][line width=0.75]    (10.93,-3.29) .. controls (6.95,-1.4) and (3.31,-0.3) .. (0,0) .. controls (3.31,0.3) and (6.95,1.4) .. (10.93,3.29)   ;
\draw    (115,34) -- (115,62) ;
\draw [shift={(115,64)}, rotate = 270] [color={rgb, 255:red, 0; green, 0; blue, 0 }  ][line width=0.75]    (10.93,-3.29) .. controls (6.95,-1.4) and (3.31,-0.3) .. (0,0) .. controls (3.31,0.3) and (6.95,1.4) .. (10.93,3.29)   ;
\draw    (155,94) -- (127,94) ;
\draw [shift={(125,94)}, rotate = 360] [color={rgb, 255:red, 0; green, 0; blue, 0 }  ][line width=0.75]    (10.93,-3.29) .. controls (6.95,-1.4) and (3.31,-0.3) .. (0,0) .. controls (3.31,0.3) and (6.95,1.4) .. (10.93,3.29)   ;
\draw    (65,94) -- (37,94) ;
\draw [shift={(35,94)}, rotate = 360] [color={rgb, 255:red, 0; green, 0; blue, 0 }  ][line width=0.75]    (10.93,-3.29) .. controls (6.95,-1.4) and (3.31,-0.3) .. (0,0) .. controls (3.31,0.3) and (6.95,1.4) .. (10.93,3.29)   ;
\draw   (215,10) -- (245,10) -- (245,40) -- (215,40) -- cycle ;
\draw   (285,10) -- (315,10) -- (315,40) -- (285,40) -- cycle ;
\draw   (355,10) -- (385,10) -- (385,40) -- (355,40) -- cycle ;
\draw    (285,25) -- (245,25) ;
\draw    (355,25) -- (315,25) ;
\draw   (250,60) -- (280,60) -- (280,90) -- (250,90) -- cycle ;
\draw   (320,60) -- (350,60) -- (350,90) -- (320,90) -- cycle ;
\draw   (390,60) -- (420,60) -- (420,90) -- (390,90) -- cycle ;
\draw    (320,75) -- (280,75) ;
\draw    (390,75) -- (350,75) ;
\draw   (215,115) -- (245,115) -- (245,145) -- (215,145) -- cycle ;
\draw   (285,115) -- (315,115) -- (315,145) -- (285,145) -- cycle ;
\draw   (355,115) -- (385,115) -- (385,145) -- (355,145) -- cycle ;
\draw    (285,130) -- (245,130) ;
\draw    (355,130) -- (315,130) ;
\draw    (230,40) -- (230,115) ;
\draw    (300,40) -- (300,115) ;
\draw    (370,40) -- (370,115) ;
\draw    (265,90) -- (240,115) ;
\draw    (335,90) -- (310,115) ;
\draw    (405,90) -- (380,115) ;
\draw    (230,145) -- (230,160) ;
\draw    (300,145) -- (300,160) ;
\draw    (370,145) -- (370,160) ;

\draw (86,154.4) node [anchor=north west][inner sep=0.75pt]    {$\sigma _{i}$};
\draw (66,16) node [anchor=north west][inner sep=0.75pt]    {$\mu _{i}$};
\draw (109,16) node [anchor=north west][inner sep=0.75pt]    {$\nu _{i}$};
\draw (86,84.4) node [anchor=north west][inner sep=0.75pt]    {$W_{i}$};
\draw (157,86) node [anchor=north west][inner sep=0.75pt]    {$r_{i+1}$};
\draw (16,86) node [anchor=north west][inner sep=0.75pt]    {$r_{i}$};
\draw (223,162.4) node [anchor=north west][inner sep=0.75pt]    {$\sigma _{1}$};
\draw (296,162.4) node [anchor=north west][inner sep=0.75pt]    {$\sigma _{2}$};
\draw (363,162.4) node [anchor=north west][inner sep=0.75pt]    {$\sigma _{3}$};
\draw (218,122.4) node [anchor=north west][inner sep=0.75pt]    {$W_{1}$};
\draw (288,122.4) node [anchor=north west][inner sep=0.75pt]    {$W_{2}$};
\draw (358,122.4) node [anchor=north west][inner sep=0.75pt]    {$W_{3}$};
\draw (256,65.4) node [anchor=north west][inner sep=0.75pt]    {$G_{1}$};
\draw (326,65.4) node [anchor=north west][inner sep=0.75pt]    {$G_{2}$};
\draw (396,65.4) node [anchor=north west][inner sep=0.75pt]    {$G_{3}$};
\draw (222,15.4) node [anchor=north west][inner sep=0.75pt]    {$F_{1}$};
\draw (292,15.4) node [anchor=north west][inner sep=0.75pt]    {$F_{2}$};
\draw (362,15.4) node [anchor=north west][inner sep=0.75pt]    {$F_{3}$};
\draw (4,7) node [anchor=north west][inner sep=0.75pt]   [align=left] {(a)};
\draw (181,7) node [anchor=north west][inner sep=0.75pt]   [align=left] {(b)};

\end{tikzpicture}

\caption{(a) The tensor cores $(W_i)^{\sigma_i}_{r_i \mu_i \nu_i r_{i+1}}$ of the convolution MPO. (b) The convolution between two functions encoded into MPSs, $F_i$ and $G_i$, is done by contracting them with the MPO $W_i$ to produce a third MPS representing the convolution. \label{fig:convolution}}
\end{center}
\end{figure}

We will follow the approach developed in Ref.~\cite{kazeev2013multilevel}, which we outline for completeness. First, rewrite the periodic convolution in Eq.~\eqref{eq:convolution} as
\begin{equation}
f_k  = \sum_{q,l \in \Omega} c_q c_l \delta^k_{ql}, \quad \delta^k_{ql} = \begin{cases}
1 & \text{if $k = q+l \mod L$ } \\
0 & \text{otherwise}
\end{cases},
\end{equation}
where we now index our Fourier space functions with integers $q,l \in \{1,2,\ldots,L\}$ instead of wave vectors for simplicity. The periodic convolution defines the three-index tensor $\delta^k_{ql}$ that encodes addition modulo $L$. We can translate the convolution into a tensor network contraction by noting that this is a contraction of two MPSs, representing the vectors $c$, with an MPO representing the tensor $\delta^{k}_{ql}$. First note that the indices $k,q,l$ are represented as bit strings in the MPS encoding. Therefore, we look for an MPO of the form
\begin{equation}
W^{\sigma}_{\mu \nu} = L^{\sigma_1}_{\mu_1 \nu_1} W^{\sigma_2}_{\mu_2 \nu_2} W^{\sigma_3}_{\mu_3 \nu_3}  \ldots R^{\sigma_N}_{\mu_N \nu_N}  = \begin{cases}
1 & \text{if $\sigma = \mu + \nu \mod L$ } \\
0  & \text{otherwise}
\end{cases},
\end{equation}
where bond indices of the MPO tensor cores are not shown explicitly and addition here refers to addition of bit strings, where modulo $L$ means modulo the binary representation of the integer $L$. In order to do this, we use the rules of addition using bit strings: suppose we have two bit strings $\mu = (\mu_1,\mu_2, \ldots , \mu_N)$ and $\nu = (\nu_1,\nu_2, \ldots , \nu_N)$ that we add together, then if we add the bits $\mu_n$ and $\nu_n$ together (not using binary addition) then we have
\begin{equation}
\mu_n + \nu_n = (\mu_n \oplus \nu_n) + R_n,
\end{equation}
where $\oplus$ is addition modulo 2 and $R_n$ is the carry bit from addition of the $n$th bits. The carry bit $R_n$ is then added to $\mu_{n-1} + \nu_{n-1}$, and so on, to construct the sum $\mu + \nu$ in the familiar way. In order to do addition modulo $L$, we simply ignore the left-most carry bit $R_1$. For example, $(1,1,1) + (0,0,1)$ has a carry bit $R_1 = 1$ which we discard, giving the sum $(0,0,0)$ which is modulo 8 as expected.

We introduce tensor cores that perform this addition process by introducing two legs for the input bits $\mu_n$ and $\nu_n$, two legs for the input and output carry bits $r_n$ and $r_{n+1}$, and one final input bit for $\sigma_n$ such that the elements of the tensor cores are given by
\begin{equation}
(W_{\mu_i \nu_i}^{\sigma_i})_{r_i r_{i+1}} = \begin{cases} 1 &  \text{if $ \mu_i \oplus \nu_i \oplus r_{i + 1} = \sigma_i$ and $ r_i = R_i $}  \\
0 & \text{otherwise}
\end{cases},
\end{equation}
where $R_i$ is the carry bit from the addition of $\mu_i$, $\nu_i$ and $r_{i+1}$. The end tensors are slightly different and do not have an outward-facing carry bit. These are
\begin{align}
L_{\mu_1 \nu_1 r_{2}}^{\sigma_1} & = \begin{cases} 
1 &  \text{if $ \mu_i \oplus \nu_i \oplus r_{i + 1} = \sigma_i$}  \\
0 & \text{otherwise}
\end{cases}, \\
R_{r_N \mu_N \nu_N}^{\sigma_1} & = \begin{cases} 
1 &  \text{if $ \mu_i \oplus \nu_i  = \sigma_i$ and $r_N = R_N$}  \\
0 & \text{otherwise}
\end{cases}.
\end{align}
Diagrammatically this is shown in Fig.~\ref{fig:convolution}(a) where the arrows represent the direction of information. The bond dimension of this tensor is $\chi_\text{MPO} = 2$ as the bond indices take values $r_i \in \{0, 1\}$. In Fig.~\ref{fig:convolution}(b) we show the contraction with two MPSs encoding a pair of functions that we convolve by contracting with the MPO.  

Once the two MPSs have been contracted with the MPO, the bond dimension of the resultant MPS will have grown considerably with bond dimension $\chi = 2\chi_f \chi_g $, where $\chi_f$ and $\chi_g$ are the bond dimensions of the MPSs encoding $f$ and $g$ respectively. This growth in bond dimension increases the size of the data, so the final step is to perform a compression by truncating the bond dimension of the resultant MPS, which can be done by performing an SVD on each tensor core similarly to how we compress in Sec.~\ref{sec:compression}. See Ref.~\cite{tensor_networks}.

By encoding the convolution in a contraction with an MPO this way, we gain a speedup. Given a pair of data vectors of length $L$ each encoded into an MPS of length $N = O(\log L)$, we perform the convolution by contracting them both with the MPO $W$. It is known that contracting an MPS of length $N$ with an MPO has time scaling complexity of $O(N \chi^2 \chi_\text{MPO}^2)$~\cite{tensor_networks} where $\chi_\text{MPO}$ is the bond dimension of the MPO, hence the convolution can be done in time $O(\log L)$ time. 

\begin{figure}
\begin{center}
\includegraphics[width = \textwidth]{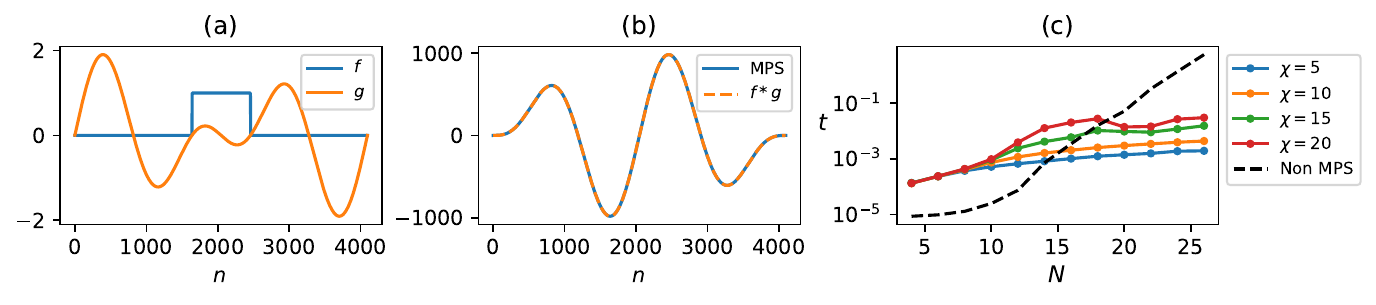}
\caption{(a) Two example periodic functions $f$ and $g$ to convolve of length $2^N$ where $N = 12$. (b) The convolution obtained via MPS-MPO contractions compared to the true convolution $f * g$, where $f$ and $g$ are both encoded into a pair of MPSs with $\chi = 3$ and $\chi = 4$ respectively (compression of $C = 0.957$ and $C = 0.928$ respectively). The resultant convolution MPS has a bond dimension of $24$ which we compress back down to $4$ giving us a compression of $C = 0.928$ again and an error of $\epsilon \approx 10^{-9}$. (c) A comparison of the time scaling of the tensor network convolution algorithm for various bond dimensions $\chi$ to the convolution done using the uncompressed raw data vectors versus the size of the data $2^N$. Raw data is convolved using $f * g = \mathfrak{Re}(\mathcal{F}^{-1}(\mathcal{F}(f)\mathcal{F}(g)))$, where $\mathcal{F}$ is numpy's fast Fourier transform. For each case we convolve the same top hat function $f$ with a random data vector $g$ whose values are sampled from $[0,1]$, and we repeat this 100 times. The time scaling for the MPS method includes the time taken to contract the MPO with the two MPSs and the time taken to compress the resultant MPS. If we do not bother compressing, the MPS scaling is better. \label{fig:convolution_example}}
\end{center}
\end{figure}

In Figs.~\ref{fig:convolution_example}(a)-(b) we show an example of convolving a top hat function with a random periodic function, and we compare to a periodic convolution of the raw data vectors in Python given by $f * g = \mathfrak{Re}(\mathcal{F}^{-1}(\mathcal{F}(f)\mathcal{F}(g)))$ where $\mathcal{F}$ is numpy's fast Fourier transform (FFT). Both functions can be encoded into a tensor network with minimal error taking a maximum bond dimension of $3$ for the top hat or the number of Fourier modes for the periodic function. We see that the tensor network implementation and exact simulation agree exactly. We also test the time scaling behaviour in Fig.~\ref{fig:convolution_example}(c) for various bond dimensions by generating random periodic data of length $2^N$ and performing the convolution with a top hat function. We see the tensor network implementation can be orders of magnitude faster than the FFT approach. As the convolution is an important spectral transformation, this complements the quantum Fourier transform of Ref.~\cite{PRXQuantum.4.040318} and will be of interest beyond CFD.

\section{Discussion and concluding remarks}\label{sec:discussion-conclusion}

In this work we have investigated compression of discretised representations of continuous functions of length $L$ into a matrix product state of length $\log(L)$. We applied the well-known method of constructing an MPS from a vector, whereby we reshape our data and perform successive singular value decompositions. As a test case, we chose to study random Fourier series as spectral methods are relevant to computational fluid dynamics. We demonstrated both numerically and analytically that if the bond dimension of the MPS is equal to the number of Fourier modes in our data, we are guaranteed a lossless encoding. Whether or not this MPS \textit{compresses} the data depends upon whether the bond dimension is below the critical threshold of $\chi_\text{crit} \approx 2^{N/2}/3$ which we established numerically. Combining this analysis gives us an understanding of whether or not random Fourier series data can be losslessly compressed or not.

We then studied whether CFD data is well-suited to the structure of an MPS. In CFD, the energy cascade states that only similar length scales interact directly. The structure of an MPS is well-suited to data with this property, as they were originally used for renormalisation techniques in quantum many-body physics. We demonstrate that solutions to the 1D Burgers' equation have minimal correlation between length scales, suggesting that MPS is a good data structure for compression. Moreover, we demonstrate that as the system evolves in time, this does not change and bond dimensions remain small allowing for good compression for thousands of time steps. 

We concluded this work with a tensor network representation of the discrete convolution with a better time scaling than if we were to perform it using the raw data vectors using the fast Fourier transform. The time scaling is promising, transforming from $O(L^2)$ with original data vectors to $O(\log L)$ when stored in MPS form.

Whilst this work was concerned with one-dimensional data, the techniques above can be simply generalised to $d$-dimensions by extending our tensors to $Nd$ in length. We also have more freedom to how data can be encoded in higher dimensions, such as different tensor network structures beyond linear MPSs including projected entangled pair states (PEPS) which have been used to study higher dimensional data in quantum many-body physics~\cite{verstraete2004renormalizationalgorithmsquantummanybody}. For multiple dimensions and lattice sizes, we may resort to alternative methods of encoding into tensors, such as tensor cross interpolation which uses machine learning methods to find the MPS structure of data from as small set of training data~\cite{10.21468/SciPostPhys.18.3.104} or by changing the ordering of physical indices $\sigma_i$ to encode different topologies of correlations beyond the length-scale encoding used here. 

The tensor network methods in this work can be translated to other hardware too. GPU versions of tensor libraries such as NVIDIA's CuQuantum and CuTensorNet contain optimised GPU versions of tensor contraction functions allowing for GPU scale ups~\cite{bayraktar2023cuquantumsdkhighperformancelibrary}. Moreover, the language of tensor networks is inherited from the language of quantum mechanics and work has been done allowing for a translation of tensor network algorithms to run on quantum computers~\cite{termanova2024tensor}, which has the potential to gain further speedup. We leave investigation of this to further work.

\section{Acknowledgment}

This research used resources of the Oak Ridge Leadership Computing Facility at Oak Ridge National Laboratory, which is supported by the Office of Science of the U.S. Department of Energy under Contract No. DE-AC05-00OR22725. The authors thank the organisers of the 2025 LEAD Symposium where this collaboration was instigated.

\bibliographystyle{elsarticle-num}
\bibliography{refs.bib}

@article{termanova2024tensor,
  title={Tensor quantum programming},
  author={Termanova, A and Melnikov, Ar and Mamenchikov, E and Belokonev, N and Dolgov, S and Berezutskii, A and Ellerbrock, R and Mansell, C and Perelshtein, MR},
  journal={New Journal of Physics},
  volume={26},
  number={12},
  pages={123019},
  year={2024},
  publisher={IOP Publishing},
  doi={10.1088/1367-2630/ad985b}
}

@article{PRXQuantum.4.040318,
  title = {Quantum Fourier Transform Has Small Entanglement},
  author = {Chen, Jielun and Stoudenmire, E.M. and White, Steven R.},
  journal = {PRX Quantum},
  volume = {4},
  issue = {4},
  pages = {040318},
  numpages = {31},
  year = {2023},
  month = {Oct},
  publisher = {American Physical Society},
  doi = {10.1103/PRXQuantum.4.040318},
  url = {https://link.aps.org/doi/10.1103/PRXQuantum.4.040318}
}

@article{doi:10.1137/090752286,
author = {Oseledets, I. V.},
title = {Tensor-Train Decomposition},
journal = {SIAM Journal on Scientific Computing},
volume = {33},
number = {5},
pages = {2295-2317},
year = {2011},
doi = {10.1137/090752286},
URL = {https://doi.org/10.1137/090752286},
eprint = {https://doi.org/10.1137/090752286}
}

@book{canuto2007spectral,
  title={Spectral Methods: Evolution to Complex Geometries and Applications to Fluid Dynamics},
  author={Canuto, C. and Hussaini, M.Y. and Quarteroni, A. and Zang, T.A.},
  isbn={9783540307280},
  lccn={2007924823},
  series={Scientific Computation},
  url={https://books.google.co.uk/books?id=7COgEw5_EBQC},
  year={2007},
  publisher={Springer Berlin Heidelberg}
}

@article{10.1098/rspa.1991.0075,
    author = {Kolmogorov, Andrei Nikolaevich and Levin, V. and Hunt, Julian Charles Roland and Phillips, Owen Martin and Williams, David},
    title = {The local structure of turbulence in incompressible viscous fluid for very large Reynolds numbers},
    journal = {Proceedings of the Royal Society of London. Series A: Mathematical and Physical Sciences},
    volume = {434},
    number = {1890},
    pages = {9-13},
    year = {1991},
    month = {07},
    issn = {0962-8444},
    doi = {10.1098/rspa.1991.0075},
    url = {https://doi.org/10.1098/rspa.1991.0075},
    eprint = {https://royalsocietypublishing.org/rspa/article-pdf/434/1890/9/68114/rspa.1991.0075.pdf},
}

@article{ali2023approximation,
  title={Approximation theory of tree tensor networks: Tensorized univariate functions},
  author={Ali, Mazen and Nouy, Anthony},
  journal={Constructive Approximation},
  volume={58},
  number={2},
  pages={463--544},
  year={2023},
  publisher={Springer},
  doi = {https://doi.org/10.1007/s00365-023-09620-w}
}

@article{RevModPhys.55.583,
  title = {The renormalization group and critical phenomena},
  author = {Wilson, Kenneth G.},
  journal = {Rev. Mod. Phys.},
  volume = {55},
  issue = {3},
  pages = {583--600},
  numpages = {0},
  year = {1983},
  month = {Jul},
  publisher = {American Physical Society},
  doi = {10.1103/RevModPhys.55.583},
  url = {https://link.aps.org/doi/10.1103/RevModPhys.55.583}
}

@Article{10.21468/SciPostPhys.18.3.104,
	title={{Learning tensor networks with tensor cross interpolation: New algorithms and libraries}},
	author={Yuriel Núñez Fernández and Marc K. Ritter and Matthieu Jeannin and Jheng-Wei Li and Thomas Kloss and Thibaud Louvet and Satoshi Terasaki and Olivier Parcollet and Jan von Delft and Hiroshi Shinaoka and Xavier Waintal},
	journal={SciPost Phys.},
	volume={18},
	pages={104},
	year={2025},
	publisher={SciPost},
	doi={10.21468/SciPostPhys.18.3.104},
	url={https://scipost.org/10.21468/SciPostPhys.18.3.104},
}

@misc{bayraktar2023cuquantumsdkhighperformancelibrary,
      title={cuQuantum SDK: A High-Performance Library for Accelerating Quantum Science}, 
      author={Harun Bayraktar and Ali Charara and David Clark and Saul Cohen and Timothy Costa and Yao-Lung L. Fang and Yang Gao and Jack Guan and John Gunnels and Azzam Haidar and Andreas Hehn and Markus Hohnerbach and Matthew Jones and Tom Lubowe and Dmitry Lyakh and Shinya Morino and Paul Springer and Sam Stanwyck and Igor Terentyev and Satya Varadhan and Jonathan Wong and Takuma Yamaguchi},
      year={2023},
      eprint={2308.01999},
      archivePrefix={arXiv},
      primaryClass={quant-ph},
      url={https://arxiv.org/abs/2308.01999}, 
}

@misc{verstraete2004renormalizationalgorithmsquantummanybody,
      title={Renormalization algorithms for Quantum-Many Body Systems in two and higher dimensions}, 
      author={F. Verstraete and J. I. Cirac},
      year={2004},
      eprint={cond-mat/0407066},
      archivePrefix={arXiv},
      primaryClass={cond-mat.str-el},
      url={https://arxiv.org/abs/cond-mat/0407066}, 
}

@inproceedings{gopalakrishnan2024towards,
  title={Towards a quantum algorithm for the incompressible nonlinear navier-stokes equations},
  author={Gopalakrishnan Meena, Muralikrishnan and Zhang, Yu and Jiang, Weiwen and Lin, Youzuo and G{\"u}nther, Stefanie and Gao, Xinfeng},
  booktitle={2024 IEEE International Conference on Quantum Computing and Engineering (QCE)},
  volume={1},
  pages={662--668},
  year={2024},
  doi = {https://doi.org/10.1109/QCE60285.2024.00083},
  organization={IEEE}
}

@article{PhysRevLett.69.2863,
  title = {Density matrix formulation for quantum renormalization groups},
  author = {White, Steven R.},
  journal = {Phys. Rev. Lett.},
  volume = {69},
  issue = {19},
  pages = {2863--2866},
  numpages = {0},
  year = {1992},
  month = {Nov},
  publisher = {American Physical Society},
  doi = {10.1103/PhysRevLett.69.2863},
  url = {https://link.aps.org/doi/10.1103/PhysRevLett.69.2863}
}

@article{PhysRevB.48.10345,
  title = {Density-matrix algorithms for quantum renormalization groups},
  author = {White, Steven R.},
  journal = {Phys. Rev. B},
  volume = {48},
  issue = {14},
  pages = {10345--10356},
  numpages = {0},
  year = {1993},
  month = {Oct},
  publisher = {American Physical Society},
  doi = {10.1103/PhysRevB.48.10345},
  url = {https://link.aps.org/doi/10.1103/PhysRevB.48.10345}
}

@inbook{gopalakrishnan2026tensor,
author = {Muralikrishnan Gopalakrishnan Meena and Vincent Jones and Yu Zhang and Xinfeng Gao},
title = {A Tensor Network-Based Quantum Algorithm for the Nonlinear 1D Burgers' Equation},
booktitle = {AIAA SCITECH 2026 Forum},
chapter = {},
pages = {},
doi = {10.2514/6.2026-1932},
URL = {https://arc.aiaa.org/doi/abs/10.2514/6.2026-1932},
eprint = {https://arc.aiaa.org/doi/pdf/10.2514/6.2026-1932}
}

@inproceedings{walters2018favre,
  title={Favre-averaged spatiotemporal-filtered large eddy simulation},
  author={Walters, Sean and Guzik, Stephen M and Gao, Xinfeng},
  booktitle={2018 AIAA Aerospace Sciences Meeting},
  pages={0373},
  year={2018},
  doi = {https://doi.org/10.2514/6.2018-0373}
}

@article{bhattacharjee26, 
title={Mechanism generating reverse buoyancy flux at the small scales of stably stratified turbulence},
volume={1026}, 
DOI={10.1017/jfm.2025.11010}, 
journal={Journal of Fluid Mechanics}, 
author={Bhattacharjee, Soumak and de Bruyn Kops, Stephen M. and Bragg, Andrew D.}, 
year={2026},
pages={A35}}

@article{pool2024nonlinear,
  title = {Nonlinear dynamics as a ground-state solution on quantum computers},
  author = {Pool, Albert J. and Somoza, Alejandro D. and Mc Keever, Conor and Lubasch, Michael and Horstmann, Birger},
  journal = {Phys. Rev. Res.},
  volume = {6},
  issue = {3},
  pages = {033257},
  numpages = {19},
  year = {2024},
  month = {Sep},
  publisher = {American Physical Society},
  doi = {10.1103/PhysRevResearch.6.033257},
  url = {https://link.aps.org/doi/10.1103/PhysRevResearch.6.033257}
}

@article{ghahremani2024deim,
title = {A DEIM Tucker tensor cross algorithm and its application to dynamical low-rank approximation},
journal = {Computer Methods in Applied Mechanics and Engineering},
volume = {423},
pages = {116879},
year = {2024},
issn = {0045-7825},
doi = {https://doi.org/10.1016/j.cma.2024.116879},
url = {https://www.sciencedirect.com/science/article/pii/S004578252400135X},
author = {Behzad Ghahremani and Hessam Babaee}
}

@article{kiffner2023tensor,
  title = {Tensor network reduced order models for wall-bounded flows},
  author = {Kiffner, Martin and Jaksch, Dieter},
  journal = {Phys. Rev. Fluids},
  volume = {8},
  issue = {12},
  pages = {124101},
  numpages = {20},
  year = {2023},
  month = {Dec},
  publisher = {American Physical Society},
  doi = {10.1103/PhysRevFluids.8.124101},
  url = {https://link.aps.org/doi/10.1103/PhysRevFluids.8.124101}
}

@article{khoromskij2010quantics,
url = {https://doi.org/10.2478/cmam-2010-0023},
title = {Quantics-TT Collocation Approximation of Parameter-Dependent and Stochastic Elliptic PDEs},
title = {},
author = {B.N. Khoromskij and I. Oseledets},
pages = {376--394},
volume = {10},
number = {4},
journal = {Computational Methods in Applied Mathematics},
doi = {doi:10.2478/cmam-2010-0023},
year = {2010},
lastchecked = {2026-04-30}
}

@article{khoromskij2011d,
  title={O (d log N)-quantics approximation of N-d tensors in high-dimensional numerical modeling},
  author={Khoromskij, Boris N},
  journal={Constructive Approximation},
  volume={34},
  number={2},
  pages={257--280},
  year={2011},
  publisher={Springer},
  doi = {https://doi.org/10.1007/s00365-011-9131-1}
}

@article{lubasch2018multigrid,
title = {Multigrid renormalization},
journal = {Journal of Computational Physics},
volume = {372},
pages = {587-602},
year = {2018},
issn = {0021-9991},
doi = {https://doi.org/10.1016/j.jcp.2018.06.065},
url = {https://www.sciencedirect.com/science/article/pii/S0021999118304431},
author = {Michael Lubasch and Pierre Moinier and Dieter Jaksch}
}

@article{garcia2021quantum,
  title={Quantum-inspired algorithms for multivariate analysis: from interpolation to partial differential equations},
  author={Garc{\'\i}a-Ripoll, Juan Jos{\'e}},
  journal={Quantum},
  volume={5},
  pages={431},
  year={2021},
  publisher={Verein zur F{\"o}rderung des Open Access Publizierens in den Quantenwissenschaften},
  url = {https://doi.org/10.22331/q-2021-04-15-431}
}

@misc{connor2025tensor-plasma,
      title={Tensor network approaches for plasma dynamics}, 
      author={Ryan J. J. Connor and Preetma Soin and Callum W. Duncan and Andrew J. Daley},
      year={2025},
      eprint={2512.15924},
      archivePrefix={arXiv},
      primaryClass={physics.plasm-ph},
      url={https://arxiv.org/abs/2512.15924}, 
}

@article{connor2025tensor-GPE,
doi = {10.1088/1367-2630/ae2b05},
url = {https://doi.org/10.1088/1367-2630/ae2b05},
year = {2026},
month = {feb},
publisher = {IOP Publishing},
volume = {28},
number = {2},
pages = {023203},
author = {Connor, Ryan J J and Duncan, Callum W and Daley, Andrew J},
title = {Tensor network methods for the Gross–Pitaevskii equation on fine grids},
journal = {New Journal of Physics}
}

@article{asztalos2024reduced,
title = {Reduced-order modeling on a near-term quantum computer},
journal = {Journal of Computational Physics},
volume = {510},
pages = {113070},
year = {2024},
issn = {0021-9991},
doi = {https://doi.org/10.1016/j.jcp.2024.113070},
url = {https://www.sciencedirect.com/science/article/pii/S002199912400319X},
author = {Katherine Asztalos and René Steijl and Romit Maulik}
}

@misc{pisoni2025compression,
      title={Compression, simulation, and synthesis of turbulent flows with tensor trains}, 
      author={Stefano Pisoni and Raghavendra Dheeraj Peddinti and Egor Tiunov and Siddhartha E. Guzman and Leandro Aolita},
      year={2026},
      eprint={2506.05477},
      archivePrefix={arXiv},
      primaryClass={physics.flu-dyn},
      url={https://arxiv.org/abs/2506.05477}, 
}

@article{gourianov2025tensor,
author = {Nikita Gourianov  and Peyman Givi  and Dieter Jaksch  and Stephen B. Pope },
title = {Tensor networks enable the calculation of turbulence probability
                    distributions},
journal = {Science Advances},
volume = {11},
number = {5},
pages = {eads5990},
year = {2025},
doi = {10.1126/sciadv.ads5990},
URL = {https://www.science.org/doi/abs/10.1126/sciadv.ads5990},
eprint = {https://www.science.org/doi/pdf/10.1126/sciadv.ads5990}}

@article{holscher2025quantum,
  title = {Quantum-inspired fluid simulation of two-dimensional turbulence with GPU acceleration},
  author = {H\"olscher, Leonhard and Rao, Pooja and M\"uller, Lukas and Klepsch, Johannes and Luckow, Andre and Stollenwerk, Tobias and Wilhelm, Frank K.},
  journal = {Phys. Rev. Res.},
  volume = {7},
  issue = {1},
  pages = {013112},
  numpages = {17},
  year = {2025},
  month = {Jan},
  publisher = {American Physical Society},
  doi = {10.1103/PhysRevResearch.7.013112},
  url = {https://link.aps.org/doi/10.1103/PhysRevResearch.7.013112}
}

@article{kazeev2013multilevel,
  title={Multilevel Toeplitz matrices generated by tensor-structured vectors and convolution with logarithmic complexity},
  author={Kazeev, Vladimir A and Khoromskij, Boris N and Tyrtyshnikov, Eugene E},
  journal={SIAM Journal on Scientific Computing},
  volume={35},
  number={3},
  pages={A1511--A1536},
  year={2013},
  doi={10.1137/110844830},
  publisher={SIAM}
}

@article{ye2022quantum,
  title = {Quantum-inspired method for solving the Vlasov-Poisson equations},
  author = {Ye, Erika and Loureiro, Nuno F. G.},
  journal = {Phys. Rev. E},
  volume = {106},
  issue = {3},
  pages = {035208},
  numpages = {20},
  year = {2022},
  month = {Sep},
  publisher = {American Physical Society},
  doi = {10.1103/PhysRevE.106.035208},
  url = {https://link.aps.org/doi/10.1103/PhysRevE.106.035208}
}

@article{gourianov2022quantum,
  title={A quantum-inspired approach to exploit turbulence structures},
  author={Gourianov, Nikita and Lubasch, Michael and Dolgov, Sergey and van den Berg, Quincy Y and Babaee, Hessam and Givi, Peyman and Kiffner, Martin and Jaksch, Dieter},
  journal={Nature Computational Science},
  volume={2},
  number={1},
  pages={30--37},
  year={2022},
  publisher={Nature Publishing Group US New York},
  doi = {https://doi.org/10.1038/s43588-021-00181-1}
}

@article{bai2019randomized,
  title={Randomized methods to characterize large-scale vortical flow networks},
  author={Bai, Zhe and Erichson, N Benjamin and Gopalakrishnan Meena, Muralikrishnan and Taira, Kunihiko and Brunton, Steven L},
  journal={PloS one},
  volume={14},
  number={11},
  pages={e0225265},
  year={2019},
  publisher={Public Library of Science San Francisco, CA USA},
  doi = {https://doi.org/10.1371/journal.pone.0225265}
}

@article{csala2022comparing,
    author = {Csala, Hunor and Dawson, Scott T. M. and Arzani, Amirhossein},
    title = {Comparing different nonlinear dimensionality reduction techniques for data-driven unsteady fluid flow modeling},
    journal = {Physics of Fluids},
    volume = {34},
    number = {11},
    pages = {117119},
    year = {2022},
    month = {11},
    issn = {1070-6631},
    doi = {10.1063/5.0127284},
    url = {https://doi.org/10.1063/5.0127284},
    eprint = {https://pubs.aip.org/aip/pof/article-pdf/doi/10.1063/5.0127284/16607728/117119_1_online.pdf},
}

@article{gopalakrishnan2018network,
  title = {Network community-based model reduction for vortical flows},
  author = {Gopalakrishnan Meena, Muralikrishnan and Nair, Aditya G. and Taira, Kunihiko},
  journal = {Phys. Rev. E},
  volume = {97},
  issue = {6},
  pages = {063103},
  numpages = {12},
  year = {2018},
  month = {Jun},
  publisher = {American Physical Society},
  doi = {10.1103/PhysRevE.97.063103},
  url = {https://link.aps.org/doi/10.1103/PhysRevE.97.063103}
}

@article{schmid2010dynamic, 
title={Dynamic mode decomposition of numerical and experimental data}, 
volume={656}, 
DOI={10.1017/S0022112010001217}, 
journal={Journal of Fluid Mechanics}, 
author={SCHMID, PETER J.}, 
year={2010}, 
pages={5–28}
}

@article{lumley1967structure,
  title={The structure of inhomogeneous turbulent flows},
  author={Lumley, John Leask},
  journal={Atmospheric turbulence and radio wave propagation},
  pages={166--178},
  year={1967},
  publisher={Nauka},
  url = {https://ci.nii.ac.jp/naid/10012381873}
}

@article{taira2017modal,
author = {Taira, Kunihiko and Brunton, Steven L. and Dawson, Scott T. M. and Rowley, Clarence W. and Colonius, Tim and McKeon, Beverley J. and Schmidt, Oliver T. and Gordeyev, Stanislav and Theofilis, Vassilios and Ukeiley, Lawrence S.},
title = {Modal Analysis of Fluid Flows: An Overview},
journal = {AIAA Journal},
volume = {55},
number = {12},
pages = {4013-4041},
year = {2017},
doi = {10.2514/1.J056060},
URL = {https://doi.org/10.2514/1.J056060},
eprint = { https://doi.org/10.2514/1.J056060}
}

@ARTICLE{manohar2018data,
  author={Manohar, Krithika and Brunton, Bingni W. and Kutz, J. Nathan and Brunton, Steven L.},
  journal={IEEE Control Systems Magazine}, 
  title={Data-Driven Sparse Sensor Placement for Reconstruction: Demonstrating the Benefits of Exploiting Known Patterns}, 
  year={2018},
  volume={38},
  number={3},
  pages={63-86},
  doi={10.1109/MCS.2018.2810460}}

@inproceedings{brewer2023entropy,
author = {Brewer, Wesley and Martinez, Daniel and Gopalakrishnan Meena, Muralikrishnan and Kashi, Aditya and Borowiec, Katarzyna and Liu, Siyan and Pilmaier, Christopher and Burgreen, Greg and Bhushan, Shanti},
title = {Entropy-driven Optimal Sub-sampling of Fluid Dynamics for Developing Machine-learned Surrogates},
year = {2023},
isbn = {9798400707858},
publisher = {Association for Computing Machinery},
address = {New York, NY, USA},
url = {https://doi.org/10.1145/3624062.3626084},
doi = {10.1145/3624062.3626084},
booktitle = {Proceedings of the SC '23 Workshops of the International Conference on High Performance Computing, Network, Storage, and Analysis},
pages = {73–80},
numpages = {8},
location = {Denver, CO, USA},
series = {SC-W '23}
}

@inproceedings{brewer2025intelligent,
author = {Brewer, Wesley and Meena Gopalakrishnan, Murali and Maiterth, Matthias and Kashi, Aditya and Choi, Jong Youl and Zhang, Pei and Nichols, Stephen and Balin, Riccardo and Couchman, Miles and de Bruyn Kops, Stephen and Yeung, P.K. and Dotson, Daniel and Uma-Vaideswaran, Rohini and Oral, Sarp and Wang, Feiyi},
title = {Intelligent Sampling of Extreme-Scale Turbulence Datasets for Accurate and Efficient Spatiotemporal Model Training},
year = {2025},
isbn = {9798400718717},
publisher = {Association for Computing Machinery},
address = {New York, NY, USA},
url = {https://doi.org/10.1145/3731599.3767340},
doi = {10.1145/3731599.3767340},
booktitle = {Proceedings of the SC '25 Workshops of the International Conference for High Performance Computing, Networking, Storage and Analysis},
pages = {1–10},
numpages = {10},
location = {
},
series = {SC Workshops '25}
}

@misc{allcock2025aurora,
      title={Aurora: Architecting Argonne's First Exascale Supercomputer for Accelerated Scientific Discovery}, 
      author={William E. Allcock and Benjamin S. Allen and James Anchell and Victor Anisimov and Thomas Applencourt and Abhishek Bagusetty and Ramesh Balakrishnan and Riccardo Balin and Solomon Bekele and Colleen Bertoni and Cyrus Blackworth and Renzo Bustamante and Kevin Canada and John Carrier and Christopher Chan-nui and Lance C. Cheney and Taylor Childers and Paul Coffman and Susan Coghlan and Tanima Dey and Michael D'Mello and Ashok Emani and Murali Emani and Kyle G. Felker and Sam Foreman and Olivier Franza and Longfei Gao and Marta García and María Garzarán and Balazs Gerofi and Yasaman Ghadar and Subrata Goswami and Neha Gupta and Kevin Harms and Väinö Hatanpää and Brian Holland and Carissa Holohan and Brian Homerding and Khalid Hossain and Xue Hu and Louise Huot and Huda Ibeid and Joseph A. Insley and Sai Jayanthi and Hong Jiang and Wei Jiang and Xiao-Yong Jin and Jeongnim Kim and Christopher Knight and Panagiotis Kourdis and Kalyan Kumaran and JaeHyuk Kwack and Janghaeng Lee and Ti Leggett and Ben Lenard and Chris Lewis and Nevin Liber and Johann Lombardi and Raymond M. Loy and Ye Luo and Bethany Lusch and Nilakantan Mahadevan and Beth Markey and Victor A. Mateevitsi and Gordon McPheeters and Ryan Milner and Jerome Mitchell and Vitali A. Morozov and Servesh Muralidharan and Tom Musta and Mrigendra Nagar and Vikram Narayana and Marieme Ngom and Anthony-Trung Nguyen and Nathan Nichols and Aditya Nishtala and James C. Osborn and Michael E. Papka and Scott Parker and Saumil S. Patel and Julia Piotrowska and Adrian C. Pope and Sucheta Raghunanda and Esteban Rangel and Paul M. Rich and Katherine M. Riley and Silvio Rizzi and Kris Rowe and Varuni Sastry and Adam Scovel and Filippo Simini and Haritha Siddabathuni Som and Patrick Steinbrecher and Rick Stevens and Xinmin Tian and Peter Upton and Thomas Uram and Archit K. Vasan and Álvaro Vázquez-Mayagoitia and Kaushik Velusamy and Brice Videau and Venkatram Vishwanath and Brian Whitney and Timothy J. Williams and Michael Woodacre and Sam Zeltner and Chuanjun Zhang and Gengbin Zheng and Huihuo Zheng},
      year={2025},
      eprint={2509.08207},
      archivePrefix={arXiv},
      primaryClass={cs.DC},
      url={https://arxiv.org/abs/2509.08207}, 
}

@INPROCEEDINGS{vazhkudai2018design,
  author={Vazhkudai, Sudharshan S. and de Supinski, Bronis R. and Bland, Arthur S. and Geist, Al and Sexton, James and Kahle, Jim and Zimmer, Christopher J. and Atchley, Scott and Oral, Sarp and Maxwell, Don E. and Larrea, Veronica G. Vergara and Bertsch, Adam and Goldstone, Robin and Joubert, Wayne and Chambreau, Chris and Appelhans, David and Blackmore, Robert and Casses, Ben and Chochia, George and Davison, Gene and Ezell, Matthew A. and Gooding, Tom and Gonsiorowski, Elsa and Grinberg, Leopold and Hanson, Bill and Hartner, Bill and Karlin, Ian and Leininger, Matthew L. and Leverman, Dustin and Marroquin, Chris and Moody, Adam and Ohmacht, Martin and Pankajakshan, Ramesh and Pizzano, Fernando and Rogers, James H. and Rosenburg, Bryan and Schmidt, Drew and Shankar, Mallikarjun and Wang, Feiyi and Watson, Py and Walkup, Bob and Weems, Lance D. and Yin, Junqi},
  booktitle={SC18: International Conference for High Performance Computing, Networking, Storage and Analysis}, 
  title={The Design, Deployment, and Evaluation of the CORAL Pre-Exascale Systems}, 
  year={2018},
  volume={},
  number={},
  pages={661-672},
  doi={10.1109/SC.2018.00055}}

@article{norman2021unprecedented,
  title={{Unprecedented cloud resolution in a GPU-enabled full-physics atmospheric climate simulation on OLCF’s summit supercomputer}},
  author={Norman, Matthew R and Bader, David A and Eldred, Christopher and Hannah, Walter M and Hillman, Benjamin R and Jones, Christopher R and Lee, Jungmin M and Leung, L R and Lyngaas, Isaac and Pressel, Kyle G and Sreepathi, Sarat and Taylor, Mark A and Yuan, Xingqiu},
  journal={International Journal of High Performance Computing Applications},
  volume={36(1)},
  pages={93--105},
  year={2021},
  doi = {https://doi.org/10.1177/10943420211027539},
  publisher={SAGE Publications Sage UK: London, England}
}

@inproceedings{atchley2023frontier,
author = {Atchley, Scott and Zimmer, Christopher and Lange, John and Bernholdt, David and Melesse Vergara, Veronica and Beck, Thomas and Brim, Michael and Budiardja, Reuben and Chandrasekaran, Sunita and Eisenbach, Markus and Evans, Thomas and Ezell, Matthew and Frontiere, Nicholas and Georgiadou, Antigoni and Glenski, Joe and Grete, Philipp and Hamilton, Steven and Holmen, John and Huebl, Axel and Jacobson, Daniel and Joubert, Wayne and Mcmahon, Kim and Merzari, Elia and Moore, Stan and Myers, Andrew and Nichols, Stephen and Oral, Sarp and Papatheodore, Thomas and Perez, Danny and Rogers, David M. and Schneider, Evan and Vay, Jean-Luc and Yeung, P. K.},
title = {Frontier: Exploring Exascale},
year = {2023},
isbn = {9798400701092},
publisher = {Association for Computing Machinery},
address = {New York, NY, USA},
url = {https://doi.org/10.1145/3581784.3607089},
doi = {10.1145/3581784.3607089},
booktitle = {Proceedings of the International Conference for High Performance Computing, Networking, Storage and Analysis},
articleno = {52},
numpages = {16},
location = {Denver, CO, USA},
series = {SC '23}
}

@article{de2019effects, 
title={The effects of stable stratification on the decay of initially isotropic homogeneous turbulence}, 
volume={860}, 
DOI={10.1017/jfm.2018.888}, 
journal={Journal of Fluid Mechanics},
author={de Bruyn Kops, Stephen M. and Riley, James J.},
year={2019}, 
pages={787–821}
}

@article{riley2023effect,
author = {James J. Riley and Miles M. P. Couchman and Stephen M. de Bruyn Kops},
title = {The effect of Prandtl number on decaying stratified turbulence},
journal = {Journal of Turbulence},
volume = {24},
number = {6-7},
pages = {330--348},
year = {2023},
publisher = {Taylor \& Francis},
doi = {10.1080/14685248.2023.2178654},
URL = {https://doi.org/10.1080/14685248.2023.2178654},
eprint = {https://doi.org/10.1080/14685248.2023.2178654}
}

@inproceedings{wilfong2025simulating,
author = {Wilfong, Benjamin and Radhakrishnan, Anand and Le Berre, Henry and Vickers, Daniel and Prathi, Tanush and Tselepidis, Nikolaos and Dorschner, Benedikt and Budiardja, Reuben and Cornille, Brian and Abbott, Stephen and Sch\"{a}fer, Florian and Bryngelson, Spencer},
title = {Simulating many-engine spacecraft: Exceeding 1 quadrillion degrees of freedom via information geometric regularization},
year = {2025},
isbn = {9798400714665},
publisher = {Association for Computing Machinery},
address = {New York, NY, USA},
url = {https://doi.org/10.1145/3712285.3771783},
doi = {10.1145/3712285.3771783},
booktitle = {Proceedings of the International Conference for High Performance Computing, Networking, Storage and Analysis},
pages = {14–24},
numpages = {11},
location = {
},
series = {SC '25}
}

@article{yeung2025gpu,
title = {GPU-enabled extreme-scale turbulence simulations: Fourier pseudo-spectral algorithms at the exascale using OpenMP offloading},
journal = {Computer Physics Communications},
volume = {306},
pages = {109364},
year = {2025},
issn = {0010-4655},
doi = {https://doi.org/10.1016/j.cpc.2024.109364},
url = {https://www.sciencedirect.com/science/article/pii/S001046552400287X},
author = {P.K. Yeung and Kiran Ravikumar and Stephen Nichols and Rohini Uma-Vaideswaran}
}

@article{yeung2025small, 
title={Small-scale properties from exascale computations of turbulence on a $\mathbf{32\,768^3}$ periodic cube}, volume={1019},
DOI={10.1017/jfm.2025.10493}, 
journal={Journal of Fluid Mechanics}, 
author={Yeung, P.K. and Ravikumar, Kiran and Uma-Vaideswaran, Rohini and Dotson, Daniel L. and Sreenivasan, Katepalli R. and Pope, Stephen B. and Meneveau, Charles and Nichols, Stephen}, 
year={2025}, 
pages={R2}}

@article{tensor_networks,
title = {The density-matrix renormalization group in the age of matrix product states},
journal = {Annals of Physics},
volume = {326},
number = {1},
pages = {96-192},
year = {2011},
note = {January 2011 Special Issue},
issn = {0003-4916},
doi = {https://doi.org/10.1016/j.aop.2010.09.012},
url = {https://www.sciencedirect.com/science/article/pii/S0003491610001752},
author = {Ulrich Schollwöck}
}

@Inbook{burgers1940,
author="Burgers, J. M.",
editor="Nieuwstadt, F. T. M.
and Steketee, J. A.",
title="Hydrodynamics. --- Application of a model system to illustrate some points of the statistical theory of free turbulence",
bookTitle="Selected Papers of J. M. Burgers",
year="1995",
publisher="Springer Netherlands",
address="Dordrecht",
pages="390--400",
isbn="978-94-011-0195-0",
doi="10.1007/978-94-011-0195-0_12",
url="https://doi.org/10.1007/978-94-011-0195-0_12"
}

\end{document}